\documentclass[12pt]{article}
\usepackage{url} 

\newcommand{\blind}{0}

\usepackage[utf8]{inputenc}
\usepackage{comment}

\usepackage{lscape} 
\usepackage{varioref}
\usepackage[linktocpage=true]{hyperref}
\usepackage{appendix}
\usepackage[style=apa,language=american,maxnames=4,minnames=3,sortcites=true]{biblatex}
\DeclareLanguageMapping{american}{american-apa}
\usepackage{multirow}
\usepackage{booktabs}
\usepackage[flushleft]{threeparttable}
\usepackage{amsmath}
\usepackage[capitalise,noabbrev]{cleveref}
\usepackage{amssymb}
\usepackage{amsthm}
\usepackage{graphicx}
\usepackage{adjustbox}
\usepackage{float}
\usepackage{soul}
\usepackage{algorithm}
\usepackage[noend]{algpseudocode}
\newcommand{\vect}[1]{{\boldsymbol{#1}}} 
\DeclareMathOperator*{\argmin}{arg\,min} 
\usepackage{xcolor}

  {
      \theoremstyle{plain} 
      \newtheorem{assumption}{Assumption}
      \newtheorem{theorem}{Theorem}
      \newtheorem{lemma}{Lemma}
  }
\begin{document}

\def\spacingset#1{\renewcommand{\baselinestretch}%
{#1}\small\normalsize} \spacingset{1}


\if0\blind
{
  \title{\bf Highest Probability Density Conformal Regions}
  \author{Max Sampson \\
    Department of Statistics and Actuarial Science, University of Iowa\\
    and \\
    Kung-Sik Chan \\
    Department of Statistics and Actuarial Science, University of Iowa}
  \maketitle
} \fi

\if1\blind
{
  \bigskip
  \bigskip
  \bigskip
  \begin{center}
    {\LARGE\bf Title}
\end{center}
  \medskip
} \fi

\bigskip
\begin{abstract}
This paper proposes a new method for finding the highest predictive density set or region, within the heteroscedastic regression framework. This framework enjoys the property that any highest predictive density set is a translation of some scalar multiple of a highest density set for the standardized regression error, with the same prediction accuracy.     The proposed method leverages this property to efficiently compute conformal prediction regions, using signed conformal inference, kernel density estimation, in conjunction with any conditional mean, and scale estimators. While most conformal prediction methods output prediction intervals, this method adapts to the target. When the target is multi-modal, the proposed method outputs an approximation of the smallest multi-modal set. When the target is uni-modal, the proposed method outputs an approximation of the smallest interval. Under mild regularity conditions, we show that these conformal prediction sets are asymptotically close to the true smallest prediction sets. Because of the conformal guarantee, even in finite sample sizes the method has guaranteed coverage. With simulations and a real data analysis we demonstrate that the proposed method is better than existing methods when the target is multi-modal, and gives similar results when the target is uni-modal. Supplementary materials, including proofs and additional images, are available online.
\end{abstract}

\noindent%
\vfill

\newpage
\spacingset{1.75} 
\section{Introduction} \label{sec:HPD}

\sloppy We consider the problem of forming the smallest prediction sets using observed predictors, $\vect{X}_{n+1}$, for an unobserved response, $Y_{n+1},$ when $n$ i.i.d. (more generally, exchangeable) copies of $((Y_1, \vect{X}_1), (Y_2, \vect{X}_2), \ldots, (Y_n, \vect{X}_n))$, with distribution $P$ are observed. Existing methods are generally computationally intensive or not adaptive to multimodal predictive distributions; see Section~\ref{sec:related_work} for a review. Here, we propose a new approach to overcome these problems by postulating that, possibly after suitable transformations,  the data follow the heteroscadastic regression model: 
\begin{equation}
    Y=g(\vect{X})+ \sigma(\vect{X})\epsilon, \label{heteroscedastic regression model}
\end{equation}
where the standardized regression error $\epsilon$ is of zero mean and independent of $\vect{X}$. For model identifiability,  $\epsilon$ may be assumed to have unit variance or unit first absolute moment, in which case $\sigma(\cdot)$ may be interpreted as the conditional standard deviation or conditional mean absolute deviation, respectively.  This framework enjoys the property that, given $\vect{X}$, the conditional  distribution of $Y$ belongs to the location-scale family generated by the distribution of $\epsilon$. Hence, any highest predictive density set of $Y$ given $\vect{X}=\vect{x}$ is a translation of some scalar multiple of a highest density set for the standardized regression error, with the same prediction accuracy.  Below, $f$ denotes the probability density function (pdf) of $\epsilon$. Leveraging this property, we aim to utilize any black-box conditional location and scale estimators along with conformal prediction to find the smallest possible prediction sets with valid finite coverage, 
\begin{equation}
\mathbb{P}(Y_{n+1} \in C(\vect{X}_{n+1})) \geq 1-\alpha.
\label{eq:marginal_coverage}
\end{equation} 

By using conformal prediction, this coverage is guaranteed \parencite{conformal_book, dis_free_pred_COPS}. Specifically, our method uses split conformal prediction, which has three steps. The first is to split the data into a training set, $\mathcal{Z}_{tr}$, and a calibration set, $\mathcal{Z}_{cal}$. The second step is to train a model on the training set. The model chosen depends on what non-conformity score one chooses to work with. For a continuous response, conditional mean and quantile regression models are common choices \parencite{split_conformal_lei_2016, conformal_book, romanocqr}. The third step is to compute the non-conformity scores on the calibration data. The final step of forming prediction sets depends on the choice of non-conformity score. Some examples of non-conformity scores and how to compute the final prediction sets are given in \textcite{conformal_book, split_conformal_lei_2016, papadopoulos_2002, romanocqr, shafer2007tutorial}. Our goal is to utilize this existing framework of conformal prediction to find smaller, more informative prediction regions that are computationally easy to compute and leverage existing regression estimators.

\subsection{Motivation and Preview for KDE-HPD}

We borrow from the ideas first given in \textcite{Linusson_2014_signed_conformal} to adjust the tail error rates independently so that we can choose the error rate in the upper and lower tails. Assuming one knew the tail error rates to minimize the prediction interval lengths, one could use this approach to find smaller prediction sets. 

Signed-conformal regression can be used to form conformal prediction intervals  that guarantee specific tail error rates in \eqref{eq:marginal_coverage} \parencite{Linusson_2014_signed_conformal}. Define the signed error non-conformity score as $V_i = Y_i - \hat{g}(\vect{X}_i) $. Then, the signed error conformal prediction region (SECPR) is given by 
\begin{equation}
C(\vect{X}_{n+1}) = [\hat{g}(\vect{X}_{n+1}) + R_{\alpha_1}(\vect{V}; \mathcal{Z}_{cal}), \hat{g}(\vect{X}_{n+1}) + Q_{1-\alpha_2}(\vect{V}; \mathcal{Z}_{cal})],
\label{eq:signed_conformal_interval}
\end{equation}
where 
\[
Q_{\delta}(\vect{V}; \mathcal{Z}_{cal}) := \lceil (\delta) (n_{cal} + 1) \rceil \text{th smallest value in }\{V_i\},
\]
\[
R_{\delta}(\vect{V}; \mathcal{Z}_{cal}) := \lceil (\delta) (n_{cal} + 1) - 1\rceil \text{th smallest value in }\{V_i\},
\]
and
\[
\alpha = \alpha_1 + \alpha_2.
\]
For example, if one wanted a conformal prediction interval with equal tailed errors, they could take $\alpha_1 = \alpha_2 = \alpha/2$. Signed-conformal regression does not necessarily need to form an interval, it can be generalized to form prediction sets. For example, one could form two disjoint intervals, one with a lower tail error rate of 0.02 and and upper tail error rate of 0.55 and one with a lower tail error rate of 0.5 and an upper tail error rate of 0.03. This would give two non-overlapping prediction intervals, one with a coverage level of 0.43 and one with a coverage level of 0.47. The set created by taking the union would have a coverage level of 0.90. 

The problem of estimating upper-level or highest density sets has been widely studied \parencite{Polonik_1995, Cuevas1997_support_estimation, Rigollet_2009_density_set_estimation, chen2016density, Samworth_2010, Lei_Wasserman_Conformal_kernel}. The goal of highest density sets is to find the smallest set for a specified coverage level, $\alpha$. It involves estimating $\{z: f(z) > \lambda^{(\alpha)} \}$ for some $\lambda^{(\alpha)} > 0$ and density, $f$, based only on samples drawn from $f$, where $\lambda^{(\alpha)}$ is chosen such that $\int_{\{z:f(z) \leq \lambda^{(\alpha)} \}} f(y) dy = \alpha$. 

Our method, the kernel density estimator for the highest predictive density (KDE-HPD) set, attempts to combine highest density set estimation with signed-conformal regression to create small prediction sets that have valid finite-sample coverage. Our method is unique in that it can describe multi-modal error terms by using a scale estimator and a marginal standardized error density estimator instead of a conditional density estimator. Existing methods either focus on prediction intervals, or require conditional density estimators. We outline existing conformal prediction methods with similar goals to ours in~\cref{sec:related_work}. Then, in~\cref{sec:kde-hpd} we outline our method. In Section~\ref{sec:theories}, we prove  that under mild conditions, our estimated set converges to this oracle set. We then compare KDE-HPD to four exisiting methods in~\cref{sec:simulations} and~\cref{sec:data analysis}. We conclude in~\cref{sec:conclusion}

\subsection{Related Work}\label{sec:related_work}

One version of split conformal prediction is conformal quantile regression (CQR). This method uses conditional quantile regression instead of conditional mean or median regression to form prediction intervals \parencite{romanocqr, cqr}. There are a few advantages of CQR compared to conformal prediction for regression that uses the absolute difference non-conformity score. One such advantage is that it takes into account the model uncertainty for certain values of $\vect{X}$, so that not all of the prediction intervals will be of the same length \parencite{romanocqr}. Let $\hat{t}_{low}(\vect{x})$ and $\hat{t}_{high}(\vect{x})$ represent the estimated lower and upper quantile regression estimates for a predictor, $\vect{x}$. The CQR non-conformity score is then, 
\[
V_i = \max\{\hat{t}_{low}(\vect{X}_i) - Y_i, Y_i - \hat{t}_{high}(\vect{X}_i)\}.
\] The prediction interval that is output is then,
\[
C(\vect{X}_{n+1}) = [\hat{t}_{low}(\vect{X}_{n+1}) - Q_{1-\alpha}(\vect{V}; \mathcal{Z}_{cal}), \hat{t}_{high}(\vect{X}_{n+1}) + Q_{1-\alpha}(\vect{V}; \mathcal{Z}_{cal})],
\]
where 
\[
Q_{1-\alpha}(\vect{V}; \mathcal{Z}_{cal}) := (1-\alpha)(1 + \frac{1}{|\mathcal{Z}_{cal}|})-\text{th empirical quantile of }\{V_i\}.
\]
and $|\mathcal{Z}_{cal}|$ is the size of the calibration set. Though CQR attempts to control conditional coverage, it does not necessarily attempt to find the smallest prediction intervals.

One method that does attempt to control conditional coverage while finding the smallest set is HPD-split. HPD-split starts with a conditional density estimator. The conformity score is then the estimated cdf evaluated at the observed responses. That is, for $(Y_i, \vect{X}_i) \in \mathcal{Z}_{cal}$, 
\[
V_i = \int_{\{z: \hat{f}(z|\vect{X}_i) \leq \hat{f}(Y_i|\vect{X}_i) \}} \hat{f}(z|\vect{X}_i)dz.
\] The final conformalized prediction set requires a grid search over $\mathcal{Y}$ and returns the set 
\[C(\vect{X}_{n+1}) = \{y: \int_{\{z: \hat{f}(z|\vect{X}_i) \leq \hat{f}(y|\vect{X}_i) \}} \hat{f}(z|\vect{X}_{n+1})dz \geq V_{\lfloor \alpha \rfloor}\},\]
where $V_{\lfloor \alpha \rfloor}$ is the $\lfloor \alpha (n_{cal} + 1) \rfloor$th smallest value of $\vect{V}$ \parencite{izbicki2021cdsplit}. 

An alternative approach is to use the density to find the smallest interval instead of the smallest set. Conformalized histogram regression (CHR) uses this approach, but replaces the conditional density estimate with a conditional histogram that is constructed with a conditional quantile or conditional density estimate. Using the conditional histogram, $T$ nested prediction intervals are formed so they have (unconformalized) coverage of $\tau_t$, for $\tau_t = t/T, t = 1, \ldots, T$. The non-conformity score is then the smallest $\tau_t$ such that the response is inside the corresponding prediction interval. The final constructed conformalized prediction interval is then the unconformalized nested prediction interval at level $\hat{\tau}$, where $\hat{\tau}$ is the $\lceil (1-\alpha) (n_{cal} + 1) \rceil$th smallest value of the non-conformity scores \parencite{CHR}.

Another approach that attempts to find the smallest interval instead of the smallest set is optimal distributional conformal prediction (DCP). Using a conditional quantile model, an estimate of the conditional cdf, $F(y, \vect{x}) = P(Y \leq y|\vect{X} = \vect{x})$, is found. Then, define 
\[\hat{Q}(\tau, \vect{x}) = \inf \{y :\hat{F}(y, \vect{x}) \geq \tau \},\] 
\[\hat{L}(x) = \min\limits_{z \in [0, \alpha]} \hat{Q}(z + 1 - \alpha, \vect{x}) - \hat{Q}(z, \vect{x}),\] and
\[
\hat{b}(\vect{x}, \alpha) = \argmin_{z \in [0, \alpha]} \hat{Q}(z + 1 - \alpha, \vect{x}) - \hat{Q}(z, \vect{x}),
\] 
all of which are estimated on the training data. $\hat{b}(\vect{x}, \alpha)$ can be thought of as an estimate for the optimal lower-bound quantile. The non-conformity scores are then,
\[
V_i = |\hat{F}(Y_i, \vect{X}_i) - \hat{b}(\vect{X}_i, \alpha) - \frac{1}{2}(1 - \alpha)|,
\]
with which a final prediction interval is constructed:
\[
\{y: |\hat{F}(y, \vect{X}_{n+1}) - \hat{b}(\vect{X}_{n+1}, \alpha) - \frac{1}{2}(1 - \alpha)| \leq Q_{1 - \alpha}(\vect{V}; \mathcal{Z}_{cal}) \}.
\]
 The set returned is always an interval because $\hat{F}(\cdot, \vect{X}_{n+1})$ is monotonic.  
\parencite{DCP}. 


\section{The Proposed Algorithm}\label{sec:kde-hpd}
Define the standardized regression error as $\epsilon:=(Y-{g}(\vect{X})) / \sigma(\vect{X})$.
The following construction relies on the fact that for a prediction set with coverage rate $1-\alpha$ for $\epsilon$, $\mathcal{C}_\epsilon$, can be transformed into a prediction set with coverage rate $1-\alpha$ for $Y$, $\mathcal{C}:={g}(\vect{X})+ 
\sigma(\vect{X})\mathcal{C}_\epsilon$. The question of finding the smallest prediction set of the preceding form naturally arises, and it is well-known that the solution is unique and obtained with $\mathcal{C}_\epsilon$ being the highest density region of $\epsilon$. 
Below, we propose a new method for finding the highest predictive density set or region using signed conformal regression \parencite{Linusson_2014_signed_conformal}. 
 
We now describe our method, an extension of signed error conformal regression that estimates the highest predictive density (HPD) set called the kernel density estimator for the HPD (KDE-HPD).  As with other split conformal prediction methods \parencite{split_conformal_lei_2016, conformal_book}, we begin by splitting our data into sets used for training and calibration. Assuming a heteroscedastic error we have two training sets indexed by $\mathcal{I}_{tr1}$ and $\mathcal{I}_{tr2}$ and a calibration set indexed by $\mathcal{I}_{cal}$. Given any point estimating function, $g$, we fit $\hat{g}$ on the first training set. If we are interested in having a point estimator that minimizes the squared error loss, we can use a conditional mean.  
\[
\hat{g} \leftarrow g(\{(Y_i, \vect{X}_i): i \in \mathcal{I}_{tr1}\}).
\]
Then, using the second training set we train a model to account for heteroscedastic errors,
\[
\hat{\sigma} \leftarrow \sigma(\{(Y_i, \hat{g}(\vect{X}_i)): i \in \mathcal{I}_{tr2}\}).
\]
Recall $\sigma$ may be interpreted as either the conditional mean absolute deviation or the conditional standard  deviation. For example, one could model the conditional mean absolute deviation by building a regression model with a response of  $|Y - \hat{g}(\vect{X})|$. See~\ref{sec:var_estimation} for more on estimating the conditional standard deviation.

Now, using the trained model, we compute non-conformity scores on the calibration set:
\[
V_{k} = (Y_k - \hat{g}(\vect{X}_k)) / \hat{\sigma}(\vect{X}_k) \text{, } \forall k \in \mathcal{I}_{cal}.
\]

Next, compute density values of the non-conformity scores, $\vect{V}=(V_1,\ldots, V_{n_{cal}})^\intercal$, using a kernel density estimator (KDE). Denote the kernel density values as $\hat{f}(\cdot)$. The choice of which kernel to use is up to the user, as they will all perform differently depending on the true, but unknown, error distribution. 

\begin{figure}[ht]
    \centering
\includegraphics[scale = 0.75]{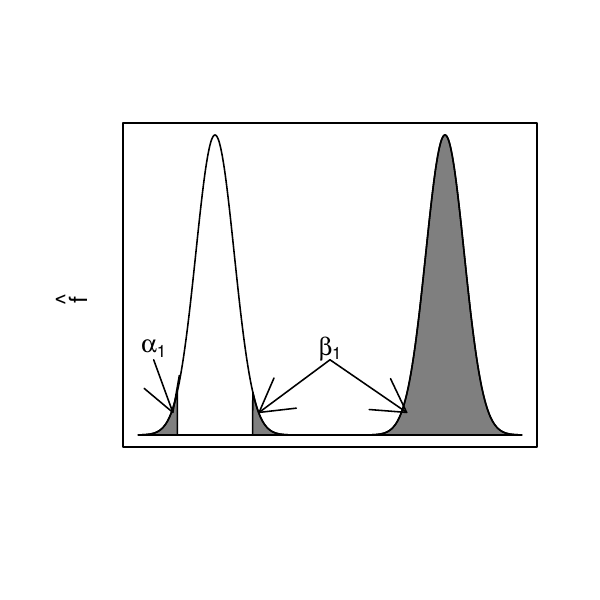}
\includegraphics[scale = 0.75]{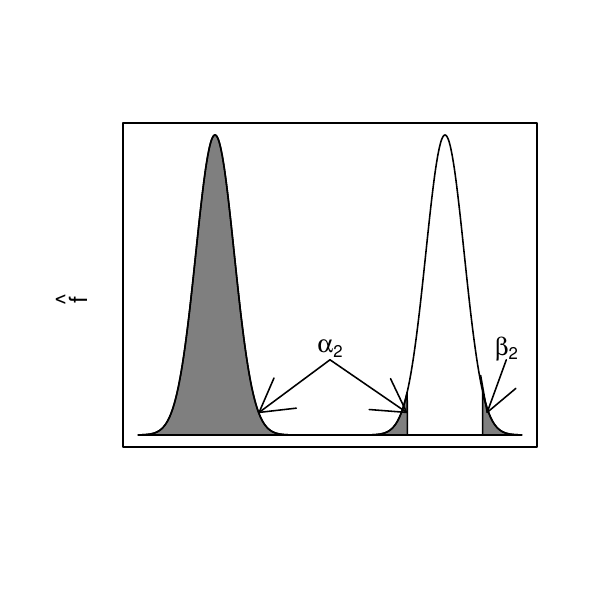}
\caption{Visualization of the upper and lower quantiles for KDE-HPD}\label{fig:kde-hpd_vis}
\end{figure}

Once we've obtained the density values, we calculate the smallest $1-\alpha$ set with $\hat{f}(\cdot)$ and the range of $\vect{V}$ \parencite{chen1999_hpd_problem, hyndman_conditional_density_1996}. We assume that the set comprises $b$ distinct intervals. Record the lower and upper endpoints of the $j-th$ interval in terms of the lower-bound quantile, ($\alpha_j$), and the upper-bound quantile, ($\beta_j$). A visualization of these quantiles for a bimodal density is given in~\cref{fig:kde-hpd_vis}. Now, find $\eta_j(\vect{X}) = R_{\alpha_j}(\vect{V}; \mathcal{Z}_{cal})$ and $\gamma_j(\vect{X}) = Q_{1-\beta_j}(\vect{V}; \mathcal{Z}_{cal})$, for $j = 1, \ldots, b$,
where 
\[
Q_{\delta}(\vect{V}; \mathcal{Z}_{cal}) := \lceil (\delta) (n_{cal} + 1) \rceil \text{th smallest value in }\{V_i\},
\]
\[
R_{\delta}(\vect{V}; \mathcal{Z}_{cal}) := \lceil (\delta) (n_{cal} + 1) - 1\rceil \text{th smallest value in }\{V_i\},
\]

Now that estimates of the quantiles for the highest predictive density set have been found, we form the interval in a similar way to \eqref{eq:signed_conformal_interval} with an adjustment for the heteroscedastic model,
\[
\hat{C}(\vect{x}) = \bigcup\limits_{j = 1}^b[\hat{g}(\vect{x}) + \eta_j(\vect{x}) \times \hat{\sigma}(\vect{x}), \hat{g}(\vect{x}) + \gamma_j(\vect{x}) \times \hat{\sigma}(\vect{x}) ].
\]
For reference, the procedure is summarized in \cref{alg:HPD_conformal}.

\begin{algorithm}
    \caption{KDE-HPD}\label{alg:HPD_conformal}
    \textbf{Input:} level $\alpha$, data = $\mathcal{Z} = (Y_i, \vect{X}_i)_{i \in \mathcal{I}}$, test point $\vect{x}$, and point estimator $g(\vect{X}; \mathcal{D})$ using $\mathcal{D}$ as data \newline
    \textbf{Procedure:}
    \begin{algorithmic}[1]

    \State Split $\mathcal{Z}$ into three folds, two training folds and one calibration fold. $\mathcal{Z}_{tr1} \overset{\Delta}{=} (Y_i, \vect{X}_i)_{i \in \mathcal{I}_{tr1}}$, $\mathcal{Z}_{tr2} \overset{\Delta}{=} (Y_i, \vect{X}_i)_{i \in \mathcal{I}_{tr2}}$, and $\mathcal{Z}_{cal} \overset{\Delta}{=} (Y_i, \vect{X}_i)_{i \in \mathcal{I}_{cal}}$
    \State Train a point estimator model, $\hat{g}(\vect{x})$, using the first training set
    \State Train a model for the conditional standard deviation, $\hat{\sigma}(\vect{x})$, using the second training set
    \State For each $i \in \mathcal{I}_{cal}$, compute the scores $V_i = (Y_i - \hat{g}(\vect{X}_i) ) / \hat{\sigma}(\vect{X}_i), \text{ for } i \in I_{cal}$
    \State Compute density values of $V_i$ using a kernel density estimator
    \State Calculate the smallest $1-\alpha$ set which is assumed to comprise $b$ distinct intervals; record the lower and upper endpoints of the $j-th$ interval in terms of the lower-bound quantile ($\alpha_j$) and the upper-bound quantile ($\beta_j$), where j = 1, 2, \ldots, b.
    \State Find $\eta_j(\vect{X}) = R_{\alpha_j}(\vect{V}; \mathcal{Z}_{cal})$ and $\gamma_j(\vect{X}) = Q_{1-\beta_j}(\vect{V}; \mathcal{Z}_{cal})$
    \end{algorithmic}
    \textbf{Output:} $\hat{C}(\vect{x}) = \bigcup\limits_{j = 1}^b[\hat{g}(\vect{x}) + \eta_j(\vect{x}) \times \hat{\sigma}(\vect{x}), \hat{g}(\vect{x}) + \gamma_j(\vect{x})\times \hat{\sigma}(\vect{x}) ]$ 

\end{algorithm}

\subsection{Heteroscedastic Model Estimation Discussion}\label{sec:var_estimation}

In our introduction of KDE-HPD, as well as in our numerical studies, we use regression with a response of $|Y - \hat{g}(\vect{X})|$ to estimate the conditional mean absolute deviation. We use this approach because it is commonly used with other conformal prediction methods, see \textcite{angelopoulos_gentle_introduction, conformal_book}. For other conditional standard deviation estimating schemes see \textcite{masry1996multivariate} for local polynomial estimation with the guarantee of uniform strong consistency, \textcite{shen_variance_estimation_2020} for a discussion on methods to estimate a conditional standard deviation function as well as their minimax convergence rates, \textcite{kolar2012variancefunctionestimationhighdimensions} for a discussion on estimating the conditional mean and conditional standard deviation in high dimensions with an $\ell_1$-norm estimator is used, and \textcite{cai_var_estimation_2009} for general estimators that do not require a conditional mean to be estimated. 

\section{Theoretical Guarantees}\label{sec:theories}

As long as our new data point comes from the same distribution as the first $n:=n_{cal}$ data points we have the standard conformal coverage guarantee, 
\begin{theorem}\label{thm:coverage_guarantee}
    \[\mathbb{P}(Y_{n+1} \in \hat{C}(\vect{X}_{n+1})) \geq 1-\alpha.\]
\end{theorem}
 This follows directly from the fact that our method is a version of the signed-conformal method given by \textcite{Linusson_2014_signed_conformal}, where we select the quantiles to attempt to minimize the length of the prediction interval. Some of these methods may be carried out with a single training sample.  

In this section when we are only discussing the kernel density estimation we denote $n_{cal}$ as $n$. Let $g(\cdot)$ and $\sigma(\cdot)$ be the chosen regression algorithm and heteroscedasticity algorithm and $\hat{g}(\cdot)$ and $\hat{\sigma}(\cdot)$ the corresponding function estimates constructed from the training data. Recall the estimated standardized regression error is $\hat{\epsilon} = (Y - \hat{g}(\vect{X}))/\hat{\sigma}(\vect{X})$. 

Assuming that $Y = g(\vect{X}) + \sigma(\vect{X})\epsilon$, prediction regions for $Y|\vect{X}$ of the form  ${g}(\vect{X})+{\sigma}(\vect{X})\times\mathcal{C}$ are the highest density prediction regions, where $\mathcal{C}$ is the highest density prediction region for ${\epsilon}$. So, instead of attempting to find the highest prediction region for $Y|\vect{X}$, we instead attempt to find the highest prediction region for $\epsilon$ and adjust it. This highest prediction region is the smallest prediction set, which is generally a union of finitely many intervals. Denote the oracle region bounds as $c_i$, $i = 1, \ldots, b$, respectively. The oracle region is then equal to $\mathcal{C}:=(c_1, c_2) \cup \ldots \cup (c_{b-1}, c_b)$, with $f(c_1) = \ldots = f(c_b)$ and common density cutoff denoted by $\lambda^{\alpha}$, which is assumed to be well-defined and unique. This value of $\lambda^{\alpha}$ is the cutoff such that $P(\{z: f(z) \leq \lambda^{\alpha} \}) = \alpha$. Let the estimated value of $\lambda^{\alpha}$ using kernel density estimation be denoted $\hat{\lambda}^{\alpha}$. Under some assumptions, we can bound how close KDE-HPD gets to the ``oracle" prediction region for $Y$ which is defined as ${g}(\vect{X}) + \sigma(\vect{X}) \times\{z: f(z)> \lambda^\alpha\} $. 

For ease of exposition, we  assume that the training sample sizes are same as $n$, the calibration sample size, although the theoretical results derived below remain the same whenever the training sample sizes are asymptotically proportional to $n$, with the limiting proportionality constants strictly bounded away from 0 and infinity. Heuristically, the proximity of $\hat{g}$ and $\hat{\sigma}$  to their population counterparts can be leveraged to study the proximity of $\{z: f(z)> \lambda^\alpha\}$ to its kernel density estimate based on the $z$'s, in terms of their Hausdorff distance. We then compare $g(\vect{x}) + \sigma(\vect{x}) \times \{z: f(z)> \lambda^\alpha\}$ to its estimate, followed by an investigation of the effects due to conformalization. The following assumptions are required for the theoretical guarantees.

\begin{assumption}\label{as:hpd1}
$(Y_i, \vect{X}_i, i=1,\ldots, n) \overset{i.i.d.}{\sim} P$ that is driven by the heteroscedastic regression model defined by (\ref{heteroscedastic regression model}). 
\end{assumption}

Let $K(z)$ be the non-negative kernel function chosen, $h$ be the bandwidth of the chosen KDE. Then, the kernel density estimate of $f(z)$ is $\hat{f}(z) = \frac{1}{nh} \sum\limits_{i = 1}^nK(\frac{z_i - z}{h})$, where $z_i=\{y_i-\hat{g}(\vect{x}_i)\}/\hat{\sigma}(\vect{x}_i)$. 

\begin{assumption}\label{as:hpd2}
$K(z)$ is symmetric about the origin, $||K||_{\infty} = K(0)$, and $\int |K(z)|^{r}  dz < \infty$ for all $r \geq 1$.
\end{assumption}

\begin{assumption}\label{as:hpd3}
$\int\limits_{-\infty}^{\infty} K(z)dx = 1$.
\end{assumption}

\begin{assumption}\label{as:hpd4}
There exist $\rho$, $C_{\rho}$, $t_0 > 0$ such that for $|t| > t_0$,
\[
K(t) \leq C_{\rho} exp(-t^{\rho}).
\]
\end{assumption}


\begin{assumption}\label{as:hpd5}
$||f||_{\infty} < M$, for some constant $M$.
\end{assumption}

\begin{assumption}\label{as:hpd6}
The density $f$ is H{\"o}lder smooth of order $\eta$ for $0 < \eta \leq 1$. That means that there exists constant $C_{\eta} > 0$ such that   $|f(z) - f(z')| \leq C_{\eta}|z - z'|^{\eta}$, $\forall z, z'$.
\end{assumption}

\begin{assumption}\label{as:hpd7}
\sloppy As $n \to \infty$, the bandwidth parameter $h \to 0$ at a rate such that  
$\log(n)/(nh) \to 0$.
\end{assumption}


\begin{assumption}\label{as:hpd8} 
Let $0 < \beta < \infty$. There exist $\lambda_0, \lambda_1$, $\check{C}_{\beta}$, $\hat{C}_{\beta}$, $\psi>0$
such that $\lambda_0<\lambda^\alpha-\psi$ so that for all $\lambda^* \in [\lambda^\alpha-\psi, \lambda^\alpha+\psi]$, the following holds for $z \in L_f(\lambda_0)\backslash L_f(\lambda^*)$.
\[
\check{C}_{\beta} \cdot d(z, L_f(\lambda^*))^{\beta} \leq \lambda^*- f(z) \leq \hat{C}_{\beta} \cdot d(z, L_f(\lambda^*))^{\beta},
\]
where $L_f(\lambda^*) := \{z: f(z) > \lambda^*\}$, $d(z, A) := \inf _{z' \in A}\{|z - z'|\}$,  $\hat{C}_{\beta}$ and $\check{C}_{\beta}$ are constants. Moreover, $L_f(\lambda_0)$ has finite Lebesgue measure. 
\end{assumption}

\begin{assumption}\label{as:hpd9} The density function $f$ satisfies the $\gamma-$exponent at level $\lambda^{\alpha}$, i.e.,  there exist constants $\tau_0 > 0$ and $b_1, b_2 > 0$, such that 
\[
b_1|\tau|^{\gamma} \leq |P(\{z: f(z) \leq \lambda^{\alpha} + \tau \}) - \alpha| \leq b_2|\tau|^{\gamma}, \> \forall \> -\tau_0 \leq \tau \leq \tau_0.
\]
\end{assumption}

\begin{assumption}\label{as:hpd10}
The conditional mean function $g(\cdot)$ is a bounded function.  The conditional variance function $\sigma^2(\cdot)$ is bounded away from zero and infinity and $\{z: {f}(z) > {\lambda}^\alpha\}$ is a bounded set. 
\end{assumption}

\begin{assumption}\label{as:hpd11}
$\hat{g}(\cdot)$ and $\hat{\sigma}(\cdot)$ are uniform strong consistent estimators, i.e., there exist deterministic sequences  $a_n\to 0, b_n\to 0$ as $n\to\infty$, such that 
\begin{eqnarray}
    &\sup_\vect{x}|\hat{g}(\vect{x}) -g(\vect{x})|=O(a_n)  \label{eqn: consistency of regression function}\\
&\sup_\vect{x}|\hat{\sigma}(\vect{x}) -\sigma(\vect{x})|=O(b_n) ,
\label{eqn: consistency of variance function}
\end{eqnarray}
almost surely, where the suprema are taken over the support of $\vect{X}$.
\end{assumption}
\begin{assumption}\label{as:hpd12}
 Let $a(\vect{x})=\hat{\sigma}^{-1}(\vect{x})\{g(\vect{x})-\hat{g}(\vect{x})\}$ and $b(\vect{x})=\hat{\sigma}^{-1}(\vect{x})\sigma(\vect{x})$. It holds that  conditional on $\hat{g}$ and $\hat{\sigma}$, 
 \begin{equation}
 \sup_z |E[b^{-1}(\vect{X})f\{b^{-1}(\vect{X}) (z - a(\vect{X})\}] -
    f(z)|=O(a_n+b_n), 
         \label{eqn: technical}
 \end{equation}
    where the supremum is taken over the support of $f$ and $a_n$ and $b_n$ are as in Assumption~\ref{as:hpd11}.
    \end{assumption}

\noindent {\bf Remark}: Assumptions~\ref{as:hpd2}--\ref{as:hpd4} are mild regularity conditions on the kernel function. Assumptions \ref{as:hpd5} and \ref{as:hpd6} are general smoothness conditions satisfied by commonly used density functions. Assumption~\ref{as:hpd7} imposes conditions on the bandwidth commonly used in the literature. Assumption~\ref{as:hpd8} is similar to a condition in \cite{jiang2017_uniformkde_convergence}.  
Assumption~\ref{as:hpd9} was first introduced by \textcite{Polonik_1995}, and was later used in many other papers on density estimation \parencite{Rigollet_2009_density_set_estimation, tsybakob_1997_density_estimation, lei2011efficient}. This assumption and assumption~\ref{as:hpd6} cannot hold at the same time unless $\gamma(\eta \wedge 1) \leq 1$ \parencite{audibert_tsybakov_gamma_exponent1, lei_robins_wasserman_2013}. This will always be true when $\gamma = 1$. The condition is a requirement that the density is not flat at $\lambda^{\alpha}$ (for stability), nor steep (for accurately selecting $\hat{\lambda}^{\alpha}$) \parencite{lei_robins_wasserman_2013}. Assumption~\ref{as:hpd10} consists of  a set of mild regularity conditions for the data generating process.  Assumption~\ref{as:hpd11}
holds for certain penalized parametric estimates for the conditional mean and variance function with $a_n$ of the form $n^{-c}$ for some $c>0$ under suitable conditions  \parencite{kolar2012variancefunctionestimationhighdimensions}. It also holds for  nonparametric conditional mean and conditional variance function estimates such as their local polynomial estimates of order $p$ with $a_n$ of the form $\left(\frac{\log n}{nh_n^d}\right)^{1/2} +h_n^{p+1}$ where $d$ is the dimension of $\vect{x}$ and $h_n$ the bandwidth, see Theorem 6 in \textcite{masry1996multivariate}.  
Under assumption~\ref{as:hpd10}, it is then straightforward to show that the conditional variance function estimate, obtained by subtracting the square of the local polynomial estimate of the conditional mean from that of the conditional second moment, is also uniform strong consistent at the rate of $O_p(a_n)$. Similarly, it can be shown that the same rate holds for the conditional standard deviation function estimate. 
Note that the  strong uniform consistency of $\hat{g}$ and $\hat{\sigma}$, at the rate of $O(a_n)$ and $O(b_n)$ respectively, and the condition that   $\sigma(\cdot)$  is bounded away from zero and $+\infty$ entail that the sup norm of $a(\cdot)$ and $b(\cdot)-1$ are $O(a_n)$ and $O(b_n)$, almost surely.
Assumption~\ref{as:hpd12}
 is a technical condition that is satisfied under assumptions~\ref{as:hpd10}--\ref{as:hpd11} and if $f$ is Lipschitz continuous and has a compact support.

We follow \textcite{jiang2017_uniformkde_convergence, chen2016density} in bounding the Hausdorff distance between two upper-level density sets. The difference in our approach is that we look to bound the distance between the true density and true cutoff with an estimated density and estimated cutoff instead of an estimated density and known cutoff. The proof of the following result is given in Section A of the Supplementary Materials.

\begin{theorem}\label{thm:hpd1}
Assume the validity of  assumptions~\ref{as:hpd1} -~\ref{as:hpd12}, with $\eta=1$ in assumption~\ref{as:hpd6},  and $\gamma=1$ in assumption~\ref{as:hpd9}. Suppose the bandwidth $h >  \log(n)/n$. For all $\varepsilon>0$,  there exists a constant $C$ such that  for $n$ sufficiently large,  the following holds with probability at least $1 - O(1/n)-\epsilon$. For all $\vect{x}$ in the support of $\vect{X}$,
\begin{eqnarray}
  && d_H(\hat{g}(\vect{x}) + \hat{\sigma}(\vect{x}) \times \{z: \hat{f}(z) > \hat{\lambda}^\alpha\}, g(\vect{x}) + \sigma(\vect{x}) \times \{z: {f}(z) > {\lambda}^\alpha\}) \nonumber \\
& <& C\times  \left\{a_n+b_n+\left(h +\sqrt{\frac{\log n}{n h}}+a_n+b_n\right)^{1/\beta}\right\}, \label{eqn: main bound}
\end{eqnarray}
\sloppy  where $a_n, b_n$ are as in assumption~\ref{as:hpd11}, $\beta$ is as in assumption~\ref{as:hpd8} and $d_H$ is the Hausdorff distance, $d_H(A, B) = \max \{ \sup\limits_{z \in A} d(z, B), \sup\limits_{y \in B} d(y, A)\}$. Here we define $d(z, A) = \inf\limits_{y \in A}\{|z - y|\}$.
\end{theorem}

\noindent \textbf{Remark:} Given $a_n$ and $b_n$, taking $h = n^{-1/3}$ optimizes the above result, but the result still holds when taking $h = n^{-1/5}$, which is the rate used to minimize the mean integrated squared error of a kernel density estimator. This allows the easy use of existing kernel density estimation packages.

Under the heteroscedastic regression model (\ref{heteroscedastic regression model}),  (\ref{eqn: main bound}) upper-bounds the distance between the estimated set and the true smallest set. So, the probability of picking a prediction set that is very different from the oracle prediction set is small, and goes to zero with increasingly large training and calibration sets. Adding the conformal adjustment to this provides us with both asymptotic and finite sample coverage guarantees. 

Theorem~\ref{thm:hpd1} implies that the set output by KDE-HPD is asymptotically close to the oracle set, as $n\to\infty$. To see this, for simplicity, we look at one cutoff point from the true standardized error term. Denote this cutoff point as $\tau_1$. Let $\alpha_1$ be the empirical CDF value for 
the estimated standardized residuals. 
It follows from the Glivenko–Cantelli theorem  that this will converge to the true CDF value of $\tau_1$ under suitable regularity condition, since $\hat{\epsilon}=\epsilon+\{b(\vect{x})-1\}\epsilon+a(\vect{x})$, where $a(\cdot)$ and $b(\cdot)$ are as defined in assumption~\ref{as:hpd12}. For instance, this is the case if assumptions \ref{as:hpd1}--\ref{as:hpd12} hold and  $\epsilon$ has a compact support. Now, the conformal adjustment quantile is the empirical quantile of $\frac{\alpha_1(n+1)}{n}$. Clearly this goes to $\alpha_1$ as $n \to \infty$. Finally, assume that the quantile function of $\epsilon$ is continuous. By \parencite[Lemma 21.2]{van1998asymptotic}, the empirical quantile will converge to the true quantile. So, for a large sample size the set output by KDE-HPD should be close to the oracle set. The benefit of this conformal adjustment is that when we have a small or medium sample size, we have conformal coverage guarantees.

Our result is similar to the optimality result for the shortest interval given in the first part of Theorem 2 of \textcite{CHR} for CHR and for the highest predictive density set in Theorem 25 of \textcite{izbicki2021cdsplit}. It is also similar to the result of Theorem 1 of \textcite{CQR_theory} for CQR and to Theorem 5 of \textcite{DCP} for optimal DCP, though our distance is the Hausdorff distance and not the Lebesgue measure of the symmetric set difference. The result for CQR was also for general quantiles, not for the shortest interval.

\section{Simulation Studies}\label{sec:simulations}
All code for the simulations and real data analysis can be found \hyperlink{https://github.com/maxsampson/KDE-HPD}{here on GitHub}. 
In this section, we demonstrate the performance of KDE-HPD compared to that of HPD-split, CHR, DCP, and CQR in five different scenarios. Each simulation scenario was run $1,000$ times with $1,000$ observed data points, 50 data points that were used for out of sample prediction, and a goal of $1 - \alpha = 0.90$ coverage. One predictor was generated, $X \sim \text{Unif}(-5, 5)$. We used the default settings for HPD-split, the default settings for CHR with a quantile forest, quantile regression with $X$ and $X^2$ as predictors for DCP, and a quantile forest with CQR \parencite{quantile_forests}. For our method, we used 50\% of the data in the training set and 50\% of the data in the calibration set. We correctly specified the conditional mean and estimated the coefficients using linear regression. In the bowtie simulation scenario, we included a model for heteroscedasticity. We used 25\% of the data to train the conditional mean model, 25\% were to train $\hat{\sigma}(X)$, a 90\% quantile random forest model with a response of $|Y - \hat{g}({X})|$, and 50\% in the calibration set. In all simulations we used a Normal kernel with default bandwidth selection in R scaled to be of the order $n_{cal}^{-1/3}$ instead of $n_{cal}^{-1/5}$. We compared the coverage, average size, and average run-time for the entire simulation to run in seconds. The computer used to run the simulations has a 9th Gen Intel i9-9900K with 8 cores up to 4.8GHz and 32GB of memory. The simulations were run using R Statistical Software version 4.3.2 except for CHR simulations, which was run using Python version 3.10.14. Simulation standard errors are given in parenthesis. The simulation setups are below. Results can be found in Tables \ref{tab:HPD1} - \ref{tab:HPD5} with the smallest set size and lowest computation time bolded. 
\begin{itemize}
    \item Unimodal and symmetric: $Y|X \sim \mathcal{N}(5 + 2X, 1)$
    \item Unimodal and skewed: $Y|X = 5 + 2X + \epsilon$, $\epsilon \sim \text{Gamma}(\text{Shape} = 7.5, \text{Rate} = 1)$
    \item Bimodal: $Y|X = 5 + 2X + \epsilon$, $\epsilon \sim p \mathcal{N}(-6, 1) + (1 - p) ( \mathcal{N}(6, 1))$, and $p \sim \text{Bernoulli}(0.5)$ 
    \item heteroscedastic: $Y|X = 5 + 2X + \epsilon|X$,  $\epsilon|X \sim \text{Gamma}(\text{Shape} = 1 + 2 |X|, \text{Rate} = 1 + 2|X|)$
    \item Bowtie: $Y|X = 5 + 2X + \epsilon|X$,  $\epsilon|X \sim \mathcal{N}(0, |X|)$, where $|X|$ is the standard deviation of the Normal distribution. 

\end{itemize}

\begin{table}[H]
\begin{center}
    \begin{tabular}{|c|c|c|c|}
        \hline
         Approach &  Coverage & Size & Computation Time   \\
         \hline
        HPD-split & 0.891 (0.001) & 3.873(0.008) & 29.44  \\
        CHR & 0.900 (0.001) & 3.930 (0.010) & 14.26  \\
        DCP & 0.902 (0.001) & \textbf{3.263} (0.005) & 0.496  \\
        CQR & 0.898 (0.001) & 3.803 (0.006) & 1.074  \\
        KDE-HPD & 0.903 (0.001) & 3.353 (0.005) & \textbf{0.005}\\
         \hline
    \end{tabular}
    \caption{Unimodal and Symmetric}
    \label{tab:HPD1}
\end{center}
\end{table}


\begin{table}[H]
\begin{center}
    \begin{tabular}{|c|c|c|c|}
        \hline
         Approach &  Coverage & Size & Computation Time   \\
         \hline
        HPD-split & 0.896 (0.001) & 10.535(0.020) & 29.59  \\
        CHR & 0.896 (0.001) & 10.170 (0.023) & 13.97  \\
        DCP & 0.906 (0.001) & \textbf{8.548} (0.012) & 0.488  \\
        CQR & 0.896 (0.001) & 10.027 (0.017) & 1.082  \\
        KDE-HPD & 0.901 (0.001) & 9.949 (0.025) & \textbf{0.005}\\
         \hline
    \end{tabular}
    \caption{Unimodal and Skewed}
    \label{tab:HPD2}
\end{center}
\end{table}


\begin{table}[H]
\begin{center}
    \begin{tabular}{|c|c|c|c|}
        \hline
         Approach &  Coverage & Size & Computation Time   \\
         \hline
        HPD-split & 0.895 (0.001) & 12.376 (0.063) & 28.36  \\
        CHR & 0.898 (0.001) & 15.205 (0.013) & 13.99  \\
        DCP & 0.904 (0.001) & 14.526 (0.006) & 0.471  \\
        CQR & 0.900 (0.001) & 15.179 (0.010) & 1.086 \\
        KDE-HPD & 0.905 (0.001) & \textbf{10.699} (0.121) & \textbf{0.006}\\
         \hline
    \end{tabular}
    \caption{Bimodal}
    \label{tab:HPD3}
\end{center}
\end{table}


\begin{table}[H]
\begin{center}
    \begin{tabular}{|c|c|c|c|}
        \hline
         Approach &  Coverage & Size & Computation Time   \\
         \hline
        HPD-split & 0.890 (0.001) & 1.685 (0.004) & 29.53  \\
        CHR & 0.900 (0.002) & 1.721 (0.006) & 14.26  \\
        DCP & 0.903 (0.001) & \textbf{1.388} (0.002) & 0.481  \\
        CQR & 0.902 (0.001) & 1.647 (0.004) & 1.084  \\

        KDE-HPD & 0.901 (0.001) & 1.779 (0.007) & \textbf{0.005}\\
         \hline
    \end{tabular}
    \caption{heteroscedastic}
    \label{tab:HPD4}
\end{center}
\end{table}



\begin{table}[H]
\begin{center}
    \begin{tabular}{|c|c|c|c|}
        \hline
         Approach &  Coverage & Size & Computation Time   \\
         \hline
        HPD-split & 0.896 (0.001) & 14.47 (0.106) & 28.01  \\
        CHR & 0.899 (0.001) & 9.268 (0.030) & 13.28  \\
        DCP & 0.903 (0.001) & \textbf{8.322} (0.026) & 0.483  \\
        CQR & 0.901 (0.001) & 9.108 (0.025) & 1.082 \\

        KDE-HPD & 0.903 (0.001) & 9.894 (0.039) & \textbf{0.306}\\
         \hline
    \end{tabular}
    \caption{Bowtie}
    \label{tab:HPD5}
\end{center}
\end{table}



\begin{figure}[ht]
    \centering
\includegraphics[scale = 0.35]{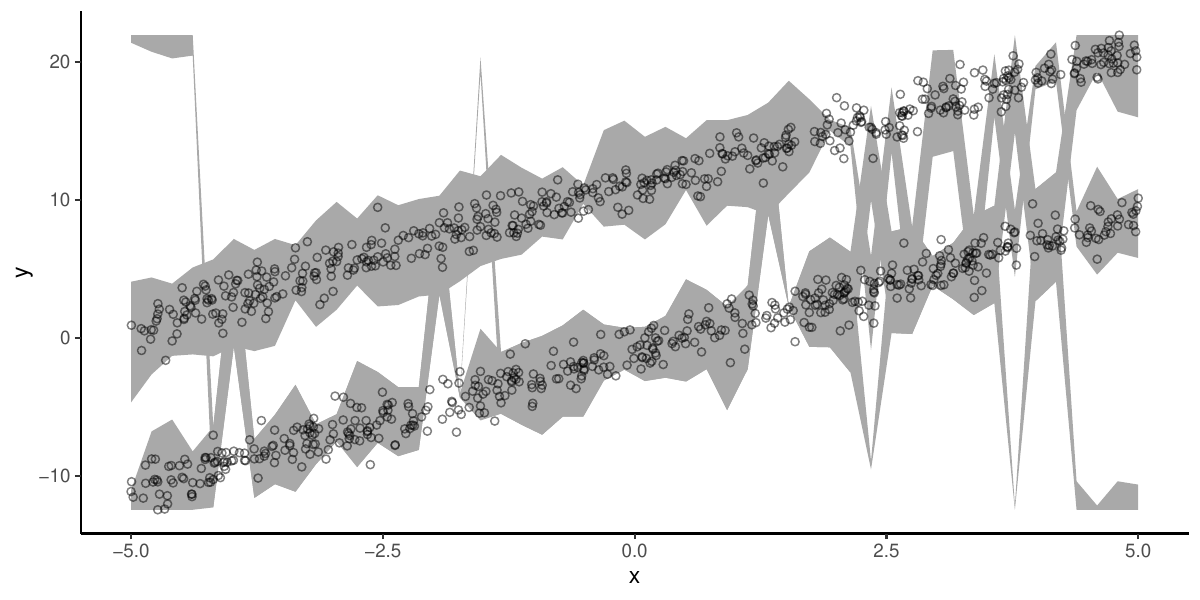}
\includegraphics[scale = 0.35]{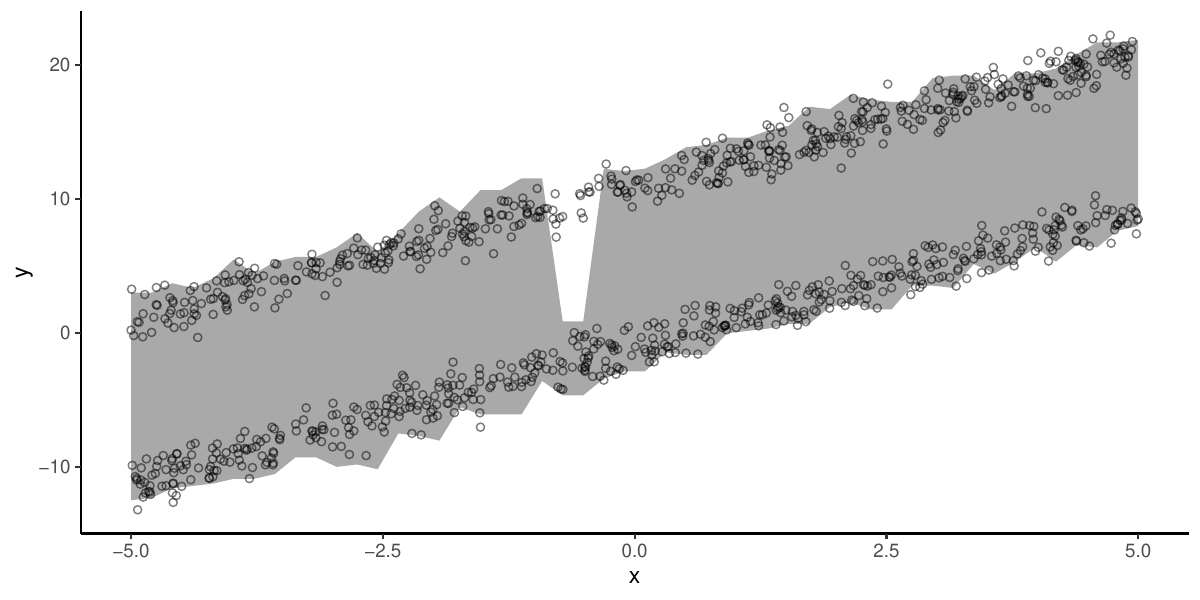}
\includegraphics[scale = 0.35]{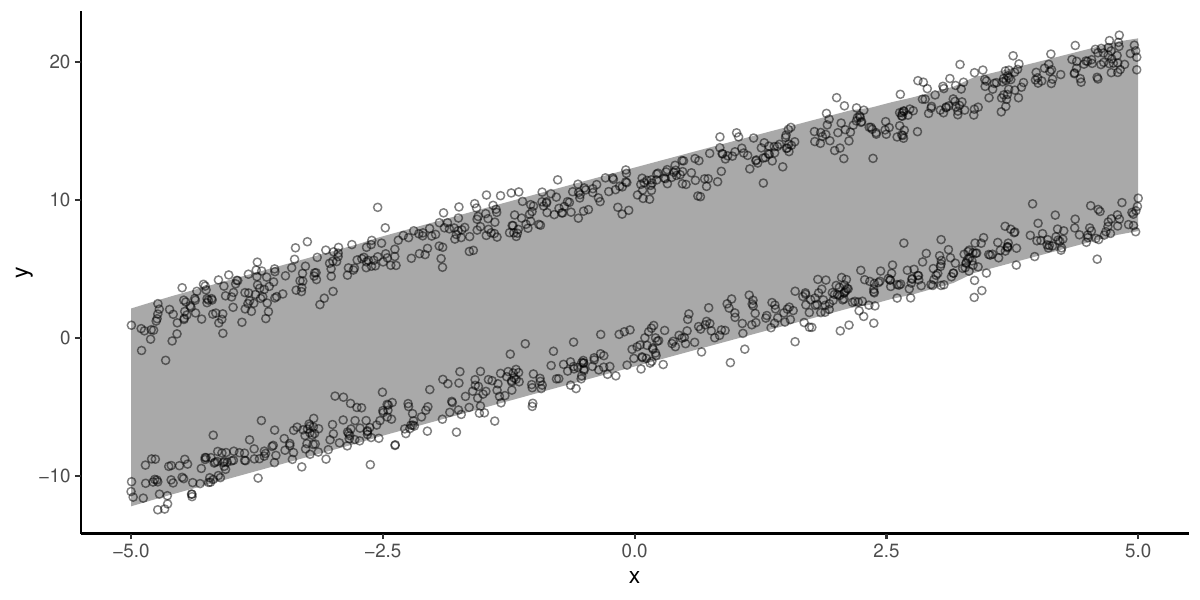}
\includegraphics[scale = 0.35]{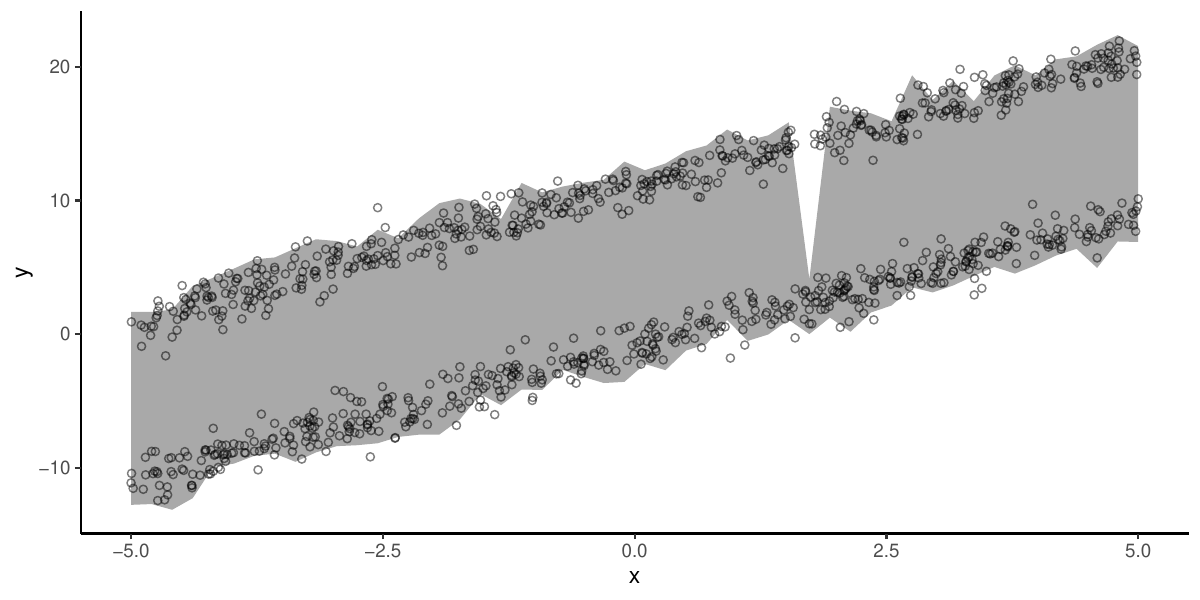}
\includegraphics[scale = 0.35]{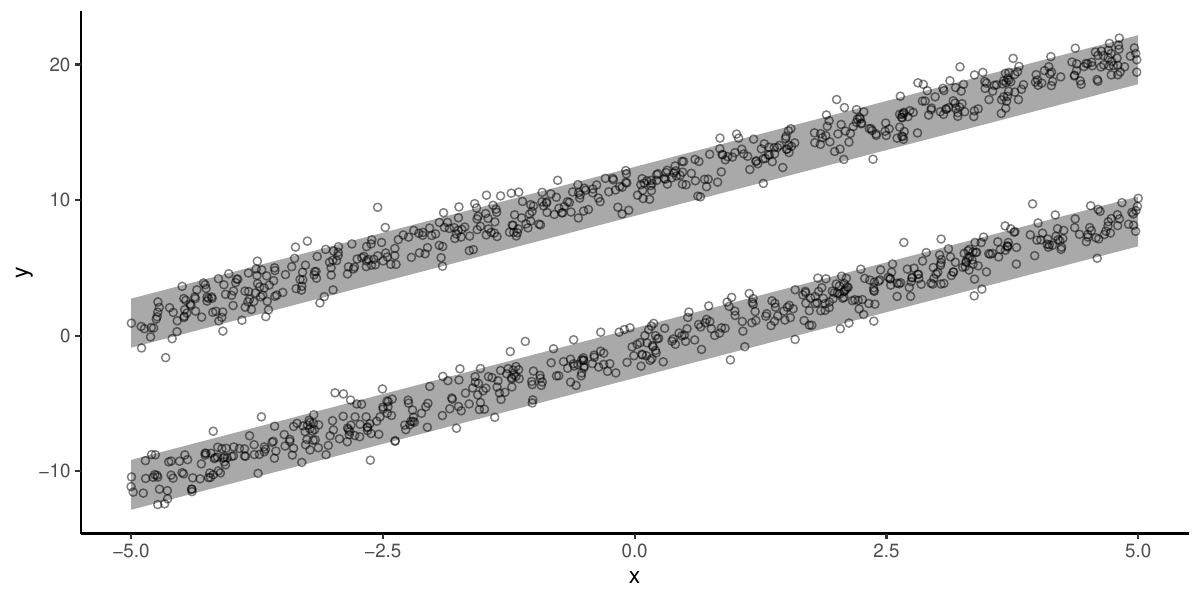}
\caption{Prediction Regions: Bimodal. The shaded region is the prediction set from one simulation. From top left to right: HPD-split, CHR, DCP, CQR, and KDE-HPD}\label{fig:pred_regions3}
\end{figure}

Examples of the prediction regions output from one simulation for the bimodal scenario and the bowtie scenario can be found in Figures~\ref{fig:pred_regions3} and~\ref{fig:pred_regions5}. The plots for the other scenarios can be found in Section C of the Supplementary Materials. We can see from the simulation results that not only is KDE-HPD tends to be faster than the other methods, especially HPD-split and CHR. It also tends to have comparable set size to the other methods, with the one exception being the bimodal error term, where KDE-HPD gives much smaller prediction regions than the competing methods. Looking at the prediction regions, we can see that when the sample size is not very large, HPD-split outputs strange regions. From~\cref{fig:pred_regions3} we can also see that CHR and CQR have spots where they fail to capture half the data. These problems are likely due to the models used, and not the conformal adjustment. The speed difference between KDE-HPD and the other methods may be small, but if one decided to use the Jackknife+ or CV+ \parencite{barber2020jackknifeplus}, the computation times would quickly add up.

\begin{figure}[ht]
    \centering
\includegraphics[scale = 0.35]{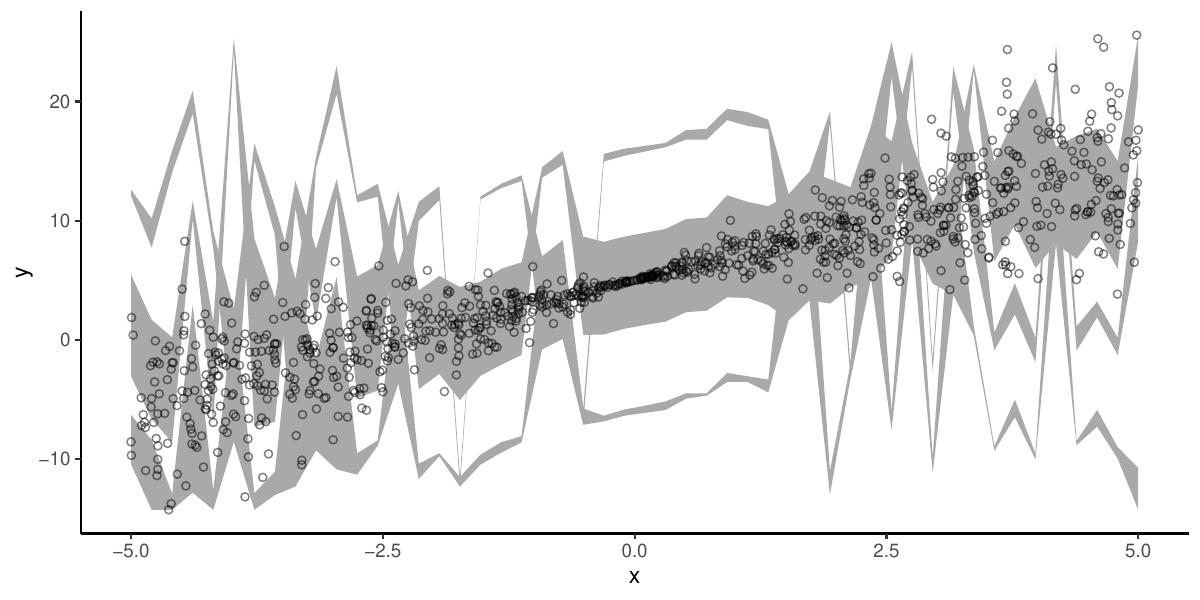}
\includegraphics[scale = 0.35]{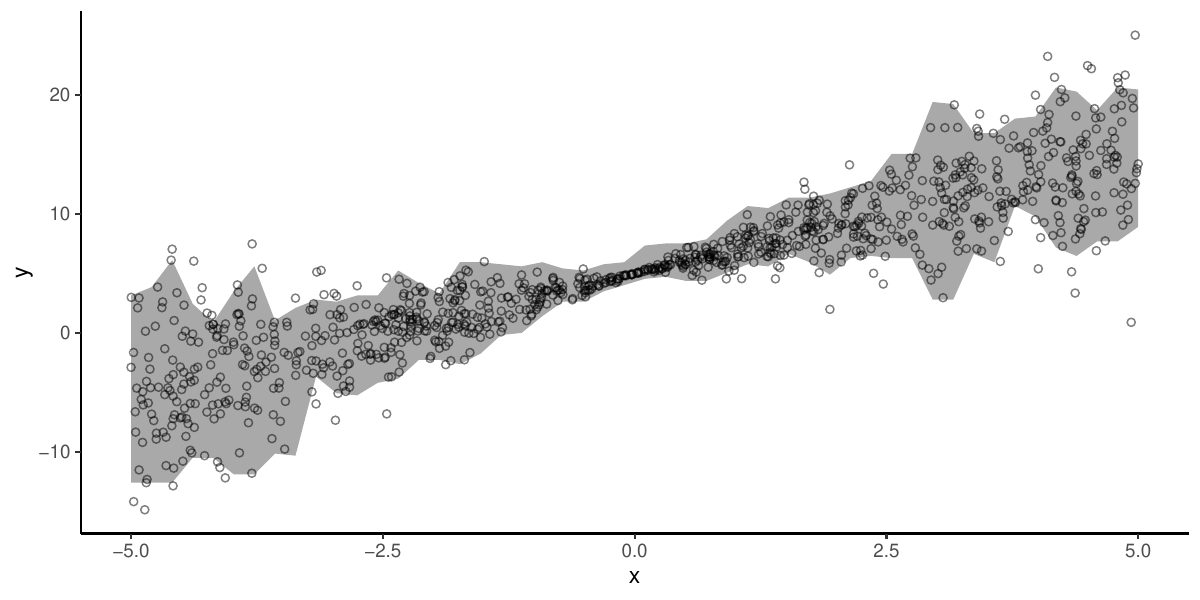}
\includegraphics[scale = 0.35]{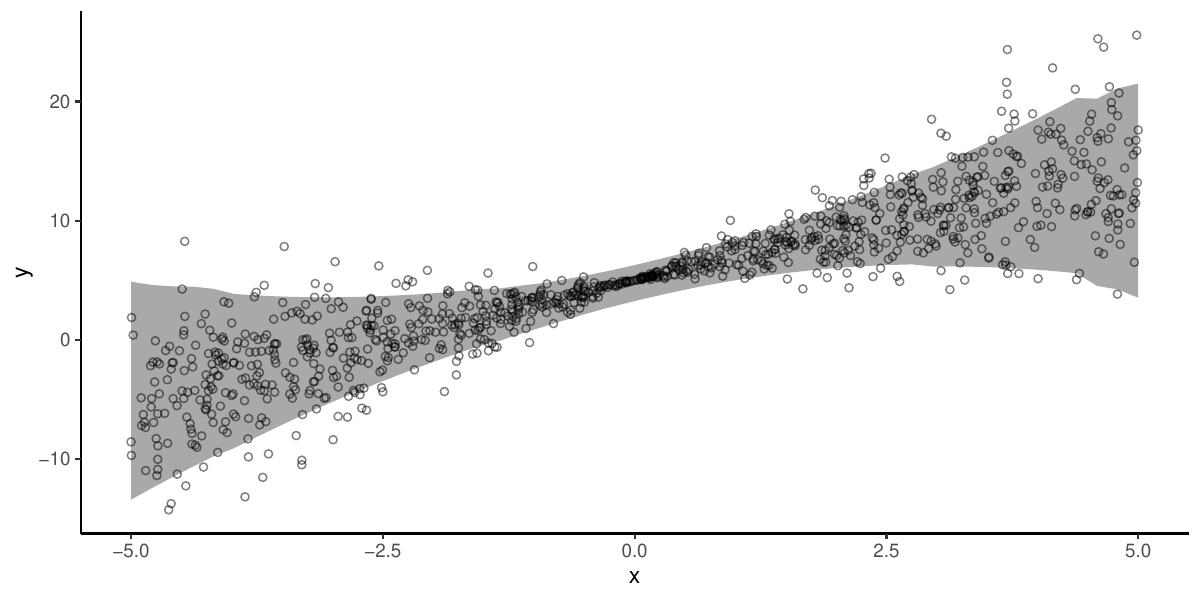}
\includegraphics[scale = 0.35]{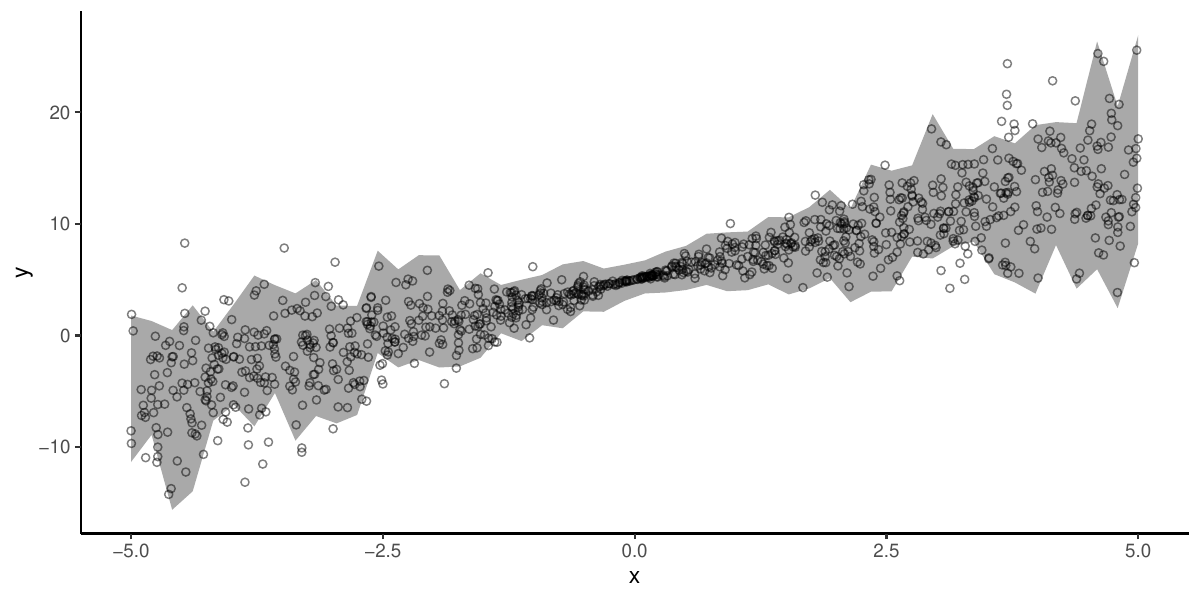}
\includegraphics[scale = 0.35]{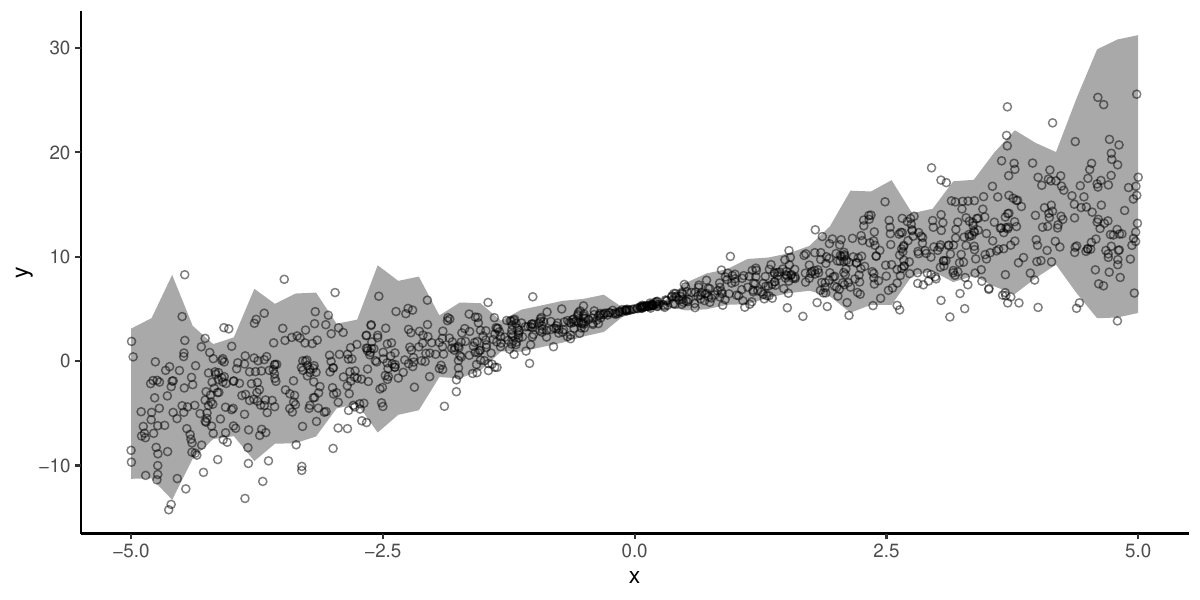}
\caption{Prediction Regions: Bowtie. The shaded region is the prediction set from one simulation. From top left to right: HPD-split, CHR, DCP, CQR, and KDE-HPD}\label{fig:pred_regions5}
\end{figure}

\section{Real Data Analysis}\label{sec:data analysis}

\begin{figure}[H]
    \centering
\includegraphics[scale = 0.5]{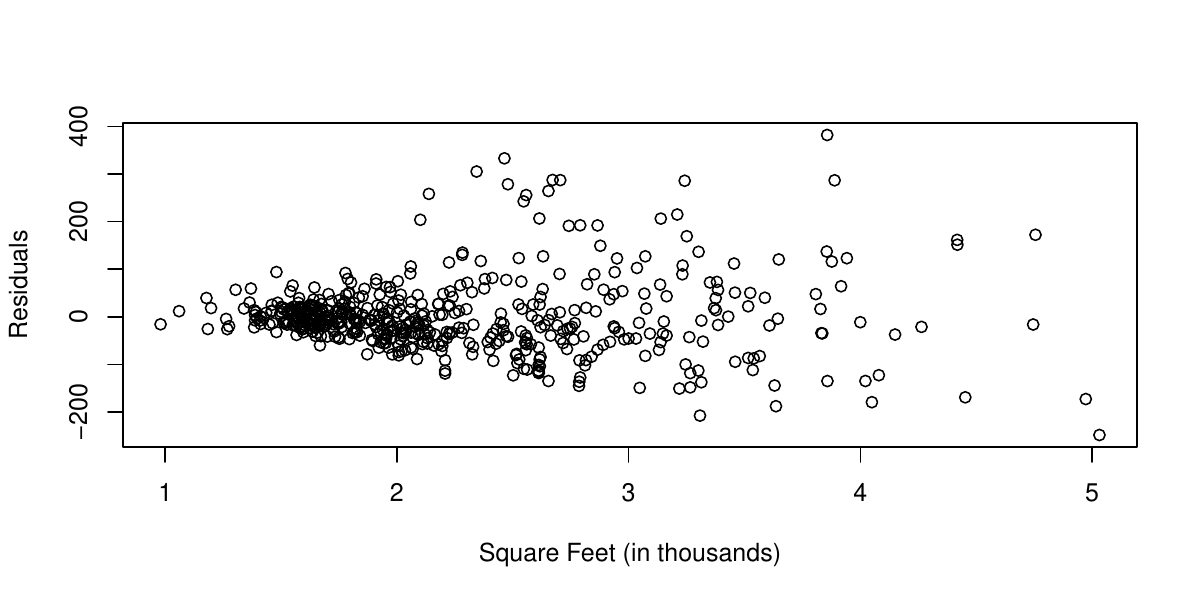}
\caption{Residuals from a linear regression vs a home's square footage}\label{fig:resid_home}
\end{figure}

A real data analysis was performed to compare KDE-HPD with HPD-split, CHR, DCP, and CQR on a data set that included the price, square footage, and air conditioning status of homes \parencite{housing_data}. We can see in~\cref{fig:resid_home}, the residuals are clearly heteroscedastic. The data were randomly permuted 200 times. There were 521 total observations, in each permutation for KDE-HPD 60 observations were used to train a linear regression for the conditional mean of the selling price, 140 were used to train a random forest for the heteroscedastic model, $\hat{\sigma} = |\text{Price} - \hat{g}(\vect{X})|$, 100 for calibration, and 221 for out of sample prediction \parencite{breiman2001random}. For all other methods, 200 observations were used to train the model, 100 were used for conformal calibration, and 221 were used for out of sample prediction. The models used were the same as the models used in the simulation studies. The average coverage, average length, and median length are given in~\cref{tab:house_results1}. Conditional coverage on AC, no AC, and selling price $> \$350, 000$ are given in~\cref{tab:house_results2}. \$350,000 is the third quartile of home prices in the data set. 

\begin{table}[H]
\begin{center}
    \begin{tabular}{|c|c|c|c|}
        \hline

         Approach & Coverage &  Mean Size & Median Size \\
         \hline
        HPD-Split & 0.878 (0.003) & 255.975 (4.137) & 212.364 (4.895)\\
        CHR & 0.893 (0.003) & 249.467 (2.218) & 204.264 (2.856) \\
        DCP & 0.907 (0.003) & 198.088 (1.616) & 171.242 (1.372)\\
        CQR & 0.905 (0.003) & 357.856 (1.538)& 362.324 (2.084)\\
        KDE-HPD & 0.906 (0.002) & 296.494 (5.458) & 230.975 (4.437) \\
        
         \hline
    \end{tabular}
    \caption{Housing Comparison Overall}
    \label{tab:house_results1}
\end{center}
\end{table}

\begin{table}[H]
\begin{center}
    \begin{tabular}{|c|c|c|c|c|c|}
        \hline

         Approach & Coverage AC & Coverage no AC & Coverage Selling Price $> \$335,000$ \\
         \hline
        HPD-Split & 0.882 (0.003) & 0.854 (0.005)& 0.680 (0.008)\\
        CHR &  0.898 (0.003) & 0.867 (0.05) & 0.822 (0.007)\\
        DCP & 0.908 (0.003) & 0.899 (0.005) & 0.788 (0.005)\\
        CQR & 0.907 (0.003) & 0.892 (0.005) & 0.874 (0.006)\\
        KDE-HPD & 0.908 (0.003)& 0.893 (0.005) & 0.874 (0.006) \\
        
         \hline
    \end{tabular}
    \caption{Housing Comparison Conditional}
    \label{tab:house_results2}
\end{center}
\end{table}

While KDE-HPD and CQR tend to have larger average lengths, it is clear from~\cref{tab:house_results2} that this is because they have better conditional coverage for expensive homes. All methods tend to do reasonably well when looking at the conditional coverage of homes with AC, though HPD-Split and CHR slightly undercover homes without AC (about 17\% of the data), leading to slightly unbalanced coverage. From this application, it is clear that KDE-HPD is adaptable to commonly found errors terms, computationally efficient, and easy to implement. 

\section{Conclusion}\label{sec:conclusion}
In this paper we introduced a new conformal prediction method, KDE-HPD, an extension of signed-conformal regression that approximates the highest predictive density set. The main benefits of KDE-HPD compared to other conformal methods are that when the target is unimodal a prediction region along with a point estimate can be given, and when the target is multi-modal a prediction region that is a union of disjoint sets can be given. 

Theoretically we show that under mild conditions the prediction regions output by KDE-HPD converge to the true smallest prediction regions. Our numerical results show that KDE-HPD performs as well as other competing methods when the target is unimodal, and much better when the target is multi-modal.

\section{Supplementary Materials}

\begin{description}
\item[Supplementary Materials:] Contains the 
 proof for Theorem~\ref{thm:hpd1}, as well as additional figures, implementation details for KDE-HPD, and another real data analysis (KDE-HPD\_Supplement.pdf).
\item[Simulation Studies and Data Analyses:] The R-code and Python-code for the data analyses and simulation studies along with the data files can be found on GitHub at https://github.com/maxsampson/KDE-HPD
\end{description}

\section{Acknowledgments}
Max Sampson was partially funded by National Institutes of Health Predoctoral Training Grant T32 HL 144461.

\newpage
\newpage
\section{References}
\printbibliography[heading=none]

\end{document}



\def\spacingset#1{\renewcommand{\baselinestretch}%
{#1}\small\normalsize} \spacingset{1}


\if0\blind
{
  \title{\bf Appendix to ``Highest Probability Density Conformal Regions” 
  }
  \author{Max Sampson \\
    Department of Statistics and Actuarial Science, University of Iowa\\
    and \\
    Kung-Sik Chan \\
    Department of Statistics and Actuarial Science, University of Iowa}
  \maketitle
} \fi

\if1\blind
{
  \bigskip
  \bigskip
  \bigskip
  \begin{center}
    {\LARGE\bf Title}
\end{center}
  \medskip
} \fi

\bigskip
\begin{abstract}
Section A contains the proof of Theorem 2  found in the manuscript. Section B contains additional figures from the simulations. Section C contains
implementation details for KDE-HPD, and Section D
contains a real data comparison of KDE-HPD with parametric prediction sets. 

\end{abstract}

\newpage
\spacingset{1.75} 
\appendix
\section{Proof of Theorem 2}\label{proof:thm1}

\begin{proof}
Recall that, for ease of exposition, we assume that the two folds of training data and the calibration data have the same sample size, denoted as $n$. 
We first bound the difference between the kernel density estimate based on  the standardized residuals $z_i=\{y_i-\hat{g}(\vect{x}_i)\}/\hat{\sigma}(\vect{x}_i)$ from that based on the true regression errors $\epsilon_i=\{y_i-g(\vect{x}_i)\}/\sigma(\vect{x}_i)$. It is readily checked that $z_i =b(\vect{x}_i)\epsilon_i+a(\vect{x}_i)$ 
where $a(\vect{x})=\hat{\sigma}^{-1}(\vect{x})\{g(\vect{x})-\hat{g}(\vect{x})\}$ and $b(\vect{x})=\hat{\sigma}^{-1}(\vect{x})\sigma(\vect{x})$.
The  strong uniform consistency of $\hat{g}$ and $\hat{\sigma}$, at the rate of $O(a_n)$ and $O(b_n)$ respectively, and the condition that   $\sigma(\cdot)$  is bounded away from zero and $+\infty$ entail that the sup norm of $a(\cdot)$ and $b(\cdot)-1$ are $O(a_n)$ and $O(b_n)$, almost surely. Thus, with no loss of generality, we may and will assume that $b(\cdot)$ is a positive function.
Conditional on $\hat{g}$ and $\hat{\sigma}$, the $z_i$'s are iid, and whose pdf at a fixed $z$ is given by (with $f(z)$ being the true pdf of $\epsilon$) 
\begin{eqnarray*}
    && f_Z(z) \\
    &=& E[b^{-1}(\vect{X})f(b^{-1}(\vect{X}) (z - a(\vect{X}))] \\
    &=& f(z)+O(a_n+b_n),
\end{eqnarray*}
thanks to assumption~12. Furthermore, it is readily seen that $f_Z(\cdot)$ is H\"older smooth with the same exponent as that of $f(\cdot)$ in assumption~6. In particular, $f_z(\cdot)$ satisfies assumptions 5 and 6.

Next, we will derive an upper  bound for  $|\lambda^{\alpha}- \hat{\lambda}^{\alpha}|$. 
\textcite[Theorem 3.3]{lei_robins_wasserman_2013} derived a related bound but they  defined $\hat{\lambda}^{\alpha}$ in terms of certain quantile of the kernel density estimate at the observed data which is different from our definition as the cutoff of an upper kernel density set, hence different proof techniques are  required. Our proof  leverages the recent result of \parencite[Theorem 2]{jiang2017_uniformkde_convergence} on the uniform convergence of the kernel density estimate to $f_Z(\cdot)$, conditional on $\hat{g}$ and $\hat{\sigma}$. Specifically, under assumptions 1--6, it holds that with probability at least $1-1/n$, uniformly in $h>(\log n/n)$,  
\begin{eqnarray}
\sup_{z\in R } |\hat{f}(z)-f_Z(z)| < C \left(h +\sqrt{\frac{\log n}{n h}}\right),
\end{eqnarray}
where  $C$ is a constant that depends only on  the exponent and the multiplicative factor in the H\"older smoothness condition, the upper bound for the density function, and the kernel function,  and $n$ is the training sample size. Therefore, conditional on $\hat{g}$ and $\hat{\sigma}$, except for an event with probability $<1/n$,  $\sup_{z\in R } |\hat{f}(z)-f(z)| < C \left(h +\sqrt{\frac{\log n}{n h}}+a_n+b_n\right)\coloneqq c_n$, with $C$  suitably enlarged if needed. 
Let $\varepsilon>0$ be a fixed constant. By further sacrificing an event of probability $\varepsilon$ to ensure the applicability of the same  (perhaps further enlarged) multiplicative factor $C$, it then holds that on an event with probability at least $1-1/n -\epsilon$, 
$\sup_{z\in R } |\hat{f}(z)-f(z)| \le c_n$ unconditionally. 
On the event $\mathcal{E}_n$ when the preceding uniform convergence holds, it is readily shown that, for any $t$, 
\begin{equation}
    \{z: f(z)\le t-c_n\} \subseteq \{z: \hat{f}(z)\le t\} \subseteq \{z: f(z)\le t+c_n\} \label{eq: subset}.
\end{equation}
Hence, 
\begin{eqnarray}
   && \int_{\{z: \hat{f}(z)\le t\}} \hat{f}(z) dz \nonumber \\
    &\le& \int_{\{z: f(z)\le t+c_n\}}  f(z)dz +\int_{\{z: f(z)\le  t+c_n\}}  \hat{f}(z)-f(z)dz\nonumber\\
   &\le& \int_{\{z: f(z)\le t+c_n\}}  f(z)dz +\int_{\{z: f(z)> t+c_n\}}  f(z)- \hat{f}(z)dz\nonumber\\
   &\le& \int_{\{z: f(z)\le t+c_n\}}  f(z)dz +c_n \times L(t+c_n) \label{eq: bound1}
\end{eqnarray}
where $L(t)$ is the Lebesgue measure of the set $\{z: f(z)> t\}$. Similarly, on $\mathcal{E}_n$, we have 
\begin{eqnarray}
     && \int_{\{z: \hat{f}(z)\le t\}} \hat{f}(z) dz \nonumber \\
     &\ge & \int_{\{z: f(z)\le t-c_n\}}  f(z)dz -c_n \times L(t-c_n). \label{eq: bound2}
\end{eqnarray}
Let $C_1>1$ be a constant to be determined.  It follows from assumption 9 and inequalities \eqref{eq: bound1} and \eqref{eq: bound2} that for all $n$ sufficiently large, 
\begin{equation}
\int_{\{z: \hat{f}(z)\le \lambda^\alpha +C_1 c_n\}} \hat{f}(z) dz \ge \alpha+b_1\times (C_1-1)\times c_n -c_n \times L(\lambda^\alpha +(C_1-1)\times c_n).
\end{equation}
Similarly, we have
\begin{equation}
\int_{\{z: \hat{f}(z)\le \lambda^\alpha -C_1 c_n\}} \hat{f}(z) dz \le \alpha-b_1\times (C_1-1)\times c_n +c_n \times L(\lambda^\alpha -(C_1-1)\times c_n).
\end{equation}
\sloppy Since $L(\cdot)$ is a decreasing function, for $n$ sufficiently large, $L(\lambda^\alpha -(C_1-1)\times c_n)<L(\lambda_0)$ which is a finite number by assumption.  By choosing $C_1$ such that $b_1(C_1-1) > 2 L(\lambda_0)$, the preceding two inequalities show that $\int_{\{z: \hat{f}(z)\le \lambda^\alpha +C_1 c_n\}}\hat{f}(z) dz>\alpha$ while $\int_{\{z: \hat{f}(z)\le \lambda^\alpha -C_1 c_n\}} \hat{f}(z) dz<\alpha$, thereby on $\mathcal{E}_n$, for $n$ sufficiently large, $|\hat{\lambda}^\alpha-\lambda^\alpha|< C_1 c_n$; furthermore, \eqref{eq: subset} entails that
\begin{equation*}
  \mathcal{Z}_1:=\{z: f(z) > \lambda^\alpha+(C_1+1)c_n\}   \subseteq \{z: \hat{f}(z) > \hat{\lambda}^\alpha\} \subseteq    
    \{z: f(z)> \lambda^\alpha-(C_1+1)c_n\}:=\mathcal{Z}_2.
\end{equation*}
Now
\begin{eqnarray*}
    && d_H(\{z: \hat{f}(z) > \hat{\lambda}^\alpha\},  \{z: f(z)> \lambda^\alpha\})\\   
    &\le& d_H(\{z: \hat{f}(z) > \hat{\lambda}^\alpha\}, \mathcal{Z}_1)+d_H(\mathcal{Z}_1, \{z: f(z)> \lambda^\alpha\})\\
     &\le& d_H(\mathcal{Z}_2, \mathcal{Z}_1)+d_H(\mathcal{Z}_1, \{z: f(z)> \lambda^\alpha\})\\
     &\le& (2(C_1+1)c_n/\check{C}_\beta)^{1/\beta} + ((C_1+1)c_n/\check{C}_\beta)^{1/\beta},
\end{eqnarray*}
owing to assumption 8. Thus, on the event $\mathcal{E}_n$ and for $n$ sufficiently large, $d_H(\{z: \hat{f}(z) > \hat{\lambda}^\alpha\},  \{z: f(z)> \lambda^\alpha\})$ is bounded by some fixed multiple of $c_n^{1/\beta}$. 

Finally, we assess the Hausdorff distance between the estimated prediction set and the oracle prediction set.
Repeated uses of the triangle inequality yield the following:
\begin{eqnarray*}
    && d_H(\hat{g}(x) + \hat{\sigma}(x) \times \{z: \hat{f}(z) > \hat{\lambda}^\alpha\}, g(x) + \sigma(x) \times \{z: {f}(z) > {\lambda}^\alpha\})  \\
    &\le& d_H(\hat{g}(x) + \sigma(x) \times \{z: {f}(z) > {\lambda}^\alpha\}, g(x) + \sigma(x) \times \{z: {f}(z) > {\lambda}^\alpha\})  \\
    &+& d_H(\hat{g}(x) + \sigma(x) \times \{z: {f}(z) > {\lambda}^\alpha\}, \hat{g}(x) + \hat{\sigma}(x) \times \{z: \hat{f}(z) > \hat{\lambda}^\alpha\})\\   
    &\le& |\hat{g}(x)- g(x)| + d_H(\hat{g}(x) + \sigma(x) \times \{z: {f}(z) > {\lambda}^\alpha\}, \hat{g}(x) + \hat{\sigma}(x) \times \{z: {f}(z) > {\lambda}^\alpha\})\\
    &+& d_H(\hat{g}(x) + \hat{\sigma}(x) \times \{z: {f}(z) > {\lambda}^\alpha\}, \hat{g}(x) + \hat{\sigma}(x) \times \{z: \hat{f}(z) > \hat{\lambda}^\alpha\})\\
    &=&  |\hat{g}(x)- g(x)| + d_H(\sigma(x) \times \{z: {f}(z) > {\lambda}^\alpha\}, \hat{\sigma}(x) \times \{z: {f}(z) > {\lambda}^\alpha\}) \\
    &+&  d_H(\hat{\sigma}(x) \times \{z: {f}(z) > {\lambda}^\alpha\}, \hat{\sigma}(x) \times \{z: \hat{f}(z) > \hat{\lambda}^\alpha\}) \\
    &\le& |\hat{g}(x)- g(x)| +
    |\hat{\sigma}(x)-\sigma(x)|\sup \{|z|: f(z)> \lambda^\alpha\}\\
    &+&
    |\hat{\sigma}(x)|\times d_H(\{z: {f}(z) > {\lambda}^\alpha\},  \{z: \hat{f}(z) > \hat{\lambda}^\alpha\})\\
    &=& O(a_n+b_n+c_n^{1/\beta}),
\end{eqnarray*}
which holds for sufficiently large $n$, 
except outside an event of probability less than $1/n+\varepsilon$
where in the last step, we have made use of assumptions 10 and 11, as well as the fact that for any sets $A, B\subset R$ and constants $a_1, c_1, a_2, c_2$,  $d_H(a_1+c_1 A, a_2+c_2 A)\le |a_1-a_2|+|c_1-c_2|\times \sup\{|a|, a\in A\}$ and $d_H(c_1 A, c_1 B)\le |c_1|d_H(A,B)$.
This completes the proof.
\end{proof}
\newpage 

\section{Conformal Regression Implementation}
A note on signed conformal regression:
One can also define the signed error non-conformity score as $V_i = \hat{g}(\vect{X}_i) - Y_i $. Then, the signed error conformal prediction region (SECPR) is given by 
\begin{equation*}
C(\vect{X}_{n+1}) = [\hat{g}(\vect{X}_{n+1}) - R_{1-\alpha_1}(\vect{V}; \mathcal{Z}_{cal}), \hat{g}(\vect{X}_{n+1}) - R_{\alpha_2}(\vect{V}; \mathcal{Z}_{cal})],
\end{equation*}
where 
\[
Q_{\delta}(\vect{V}; \mathcal{Z}_{cal}) := \lceil (\delta) (n_{cal} + 1) \rceil \text{th smallest value in }\{V_i\},
\]
\[
R_{\delta}(\vect{V}; \mathcal{Z}_{cal}) := \lceil (\delta) (n_{cal} + 1) - 1\rceil \text{th smallest value in }\{V_i\},
\]
Both approaches provide the same coverage guarantees. We provide this note because it is how we chose to code KDE-HPD. 

\newpage

\section{Additional Prediction Region Figures}

\begin{figure}[H]
    \centering
\includegraphics[scale = 0.35]{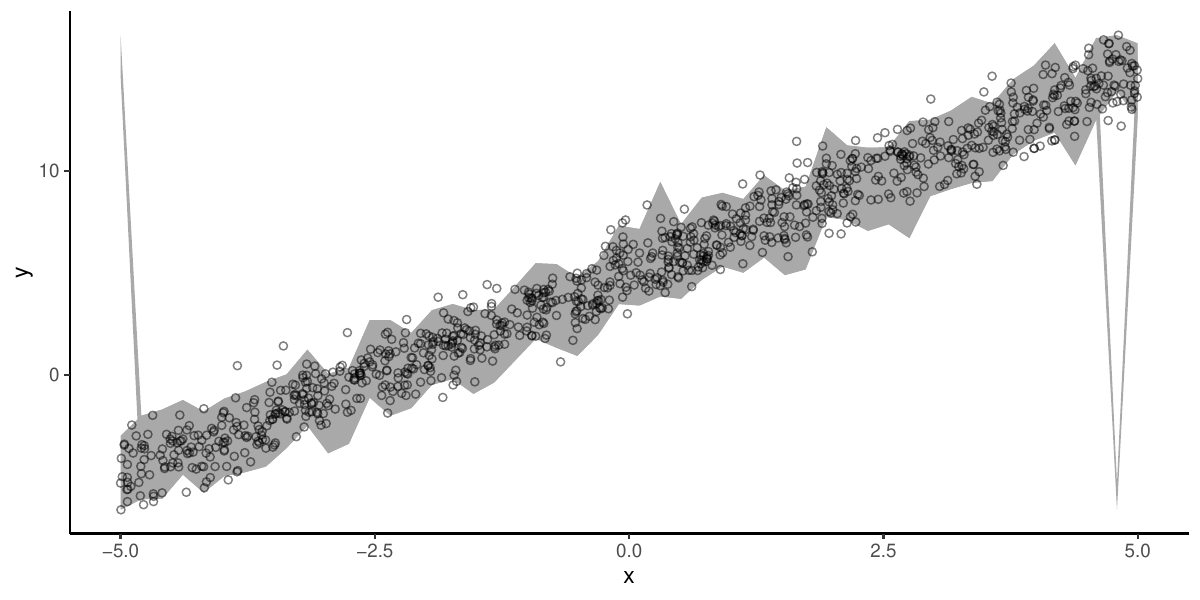}
\includegraphics[scale = 0.35]{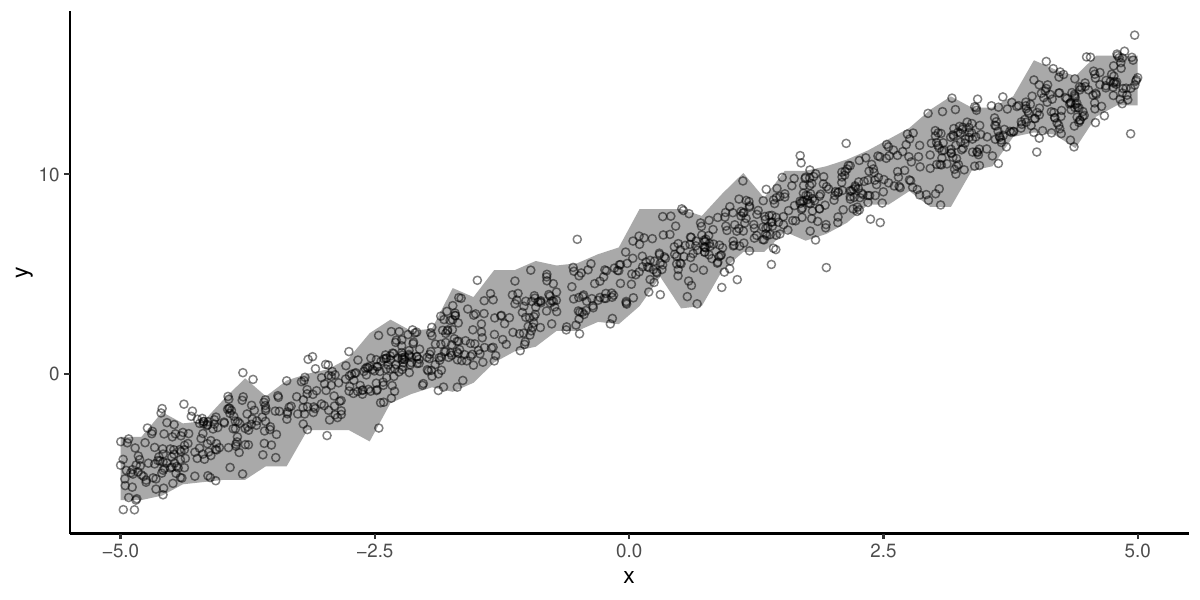}
\includegraphics[scale = 0.35]{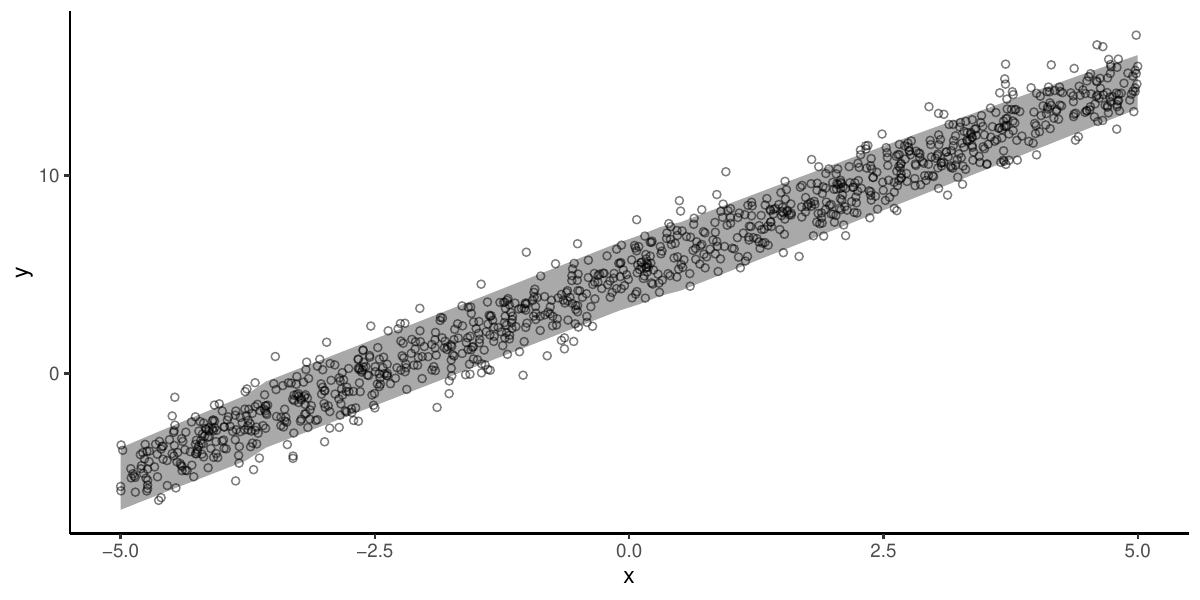}
\includegraphics[scale = 0.35]{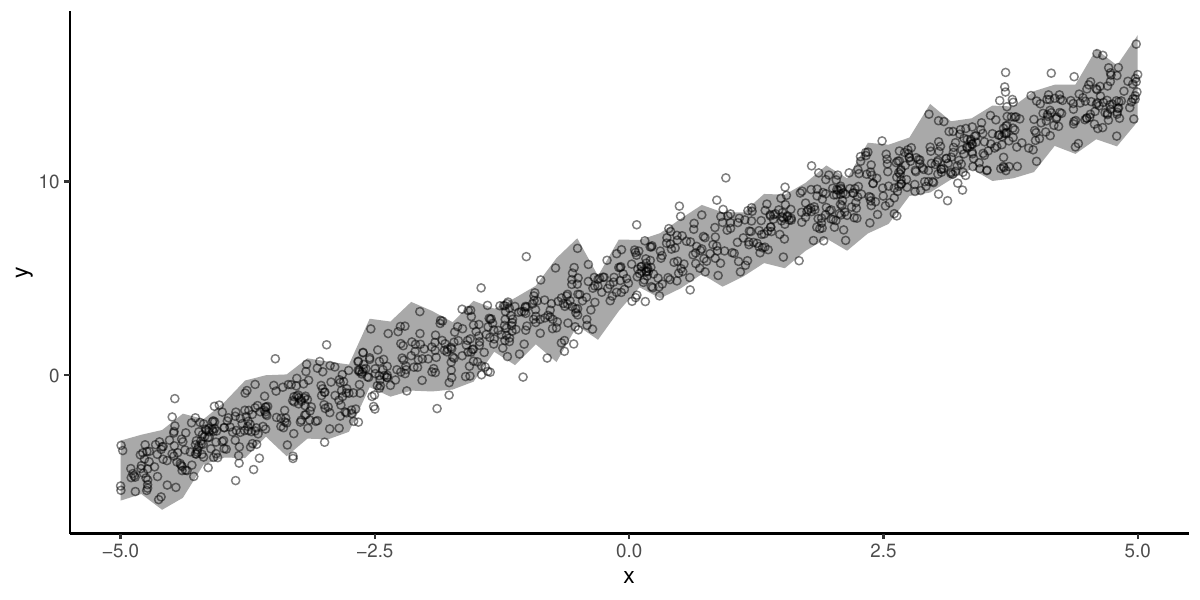}
\includegraphics[scale = 0.35]{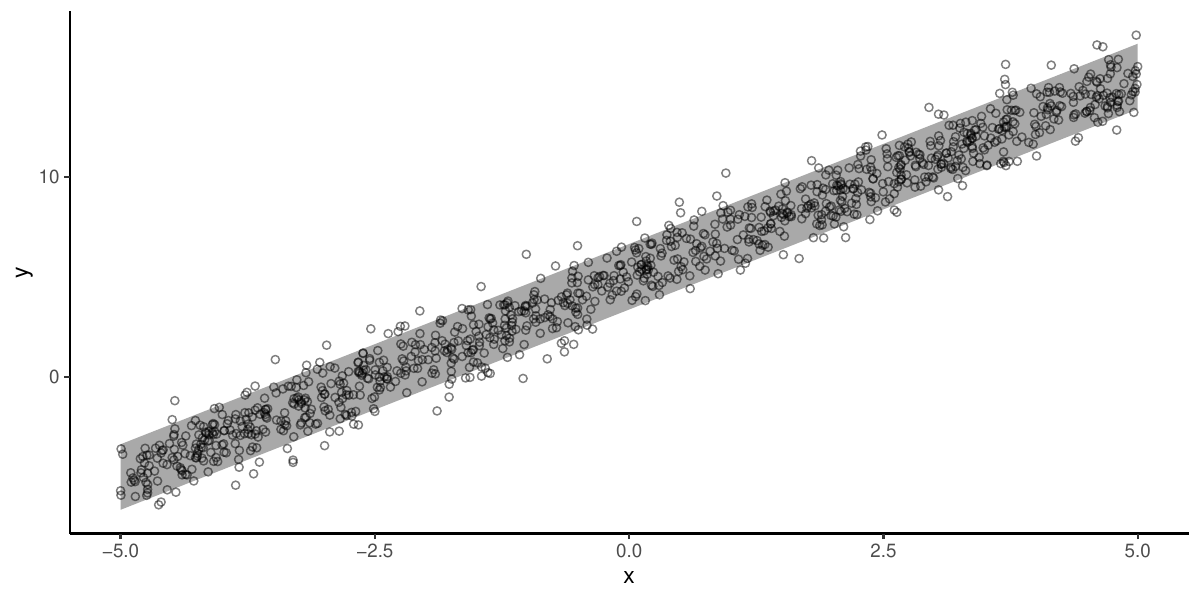}
\caption{Prediction Regions: Unimodal and Symmetric. The shaded region is the prediction set from one simulation. From top left to right: HPD-split, CHR, DCP, CQR, and KDE-HPD}\label{fig:pred_regions1}
\end{figure}

\begin{figure}[H]
    \centering
\includegraphics[scale = 0.35]{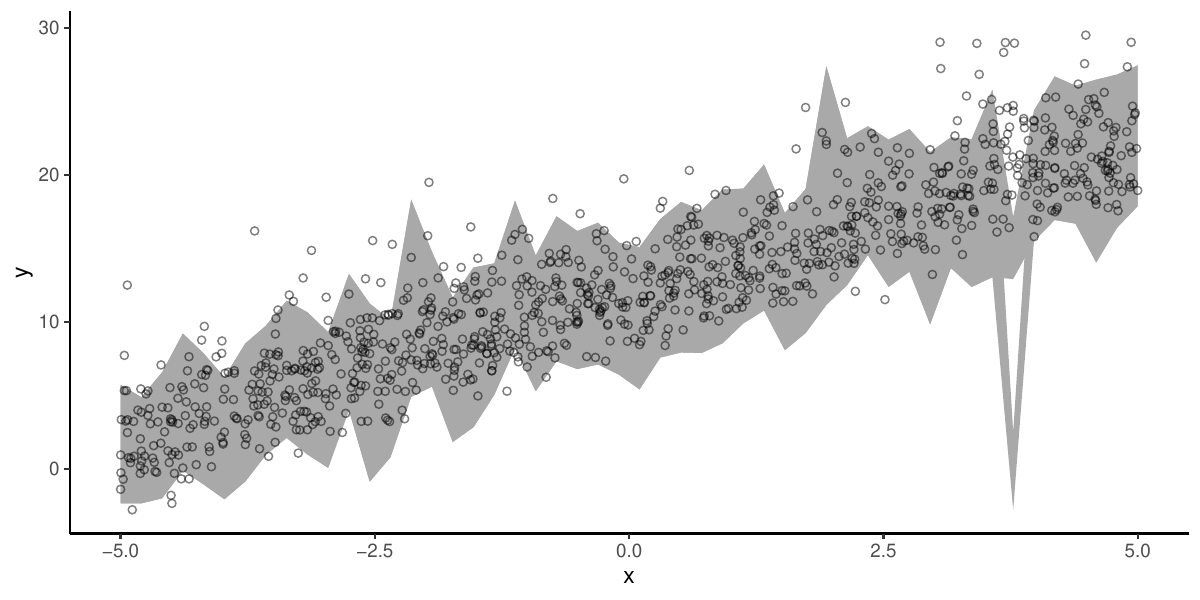}
\includegraphics[scale = 0.35]{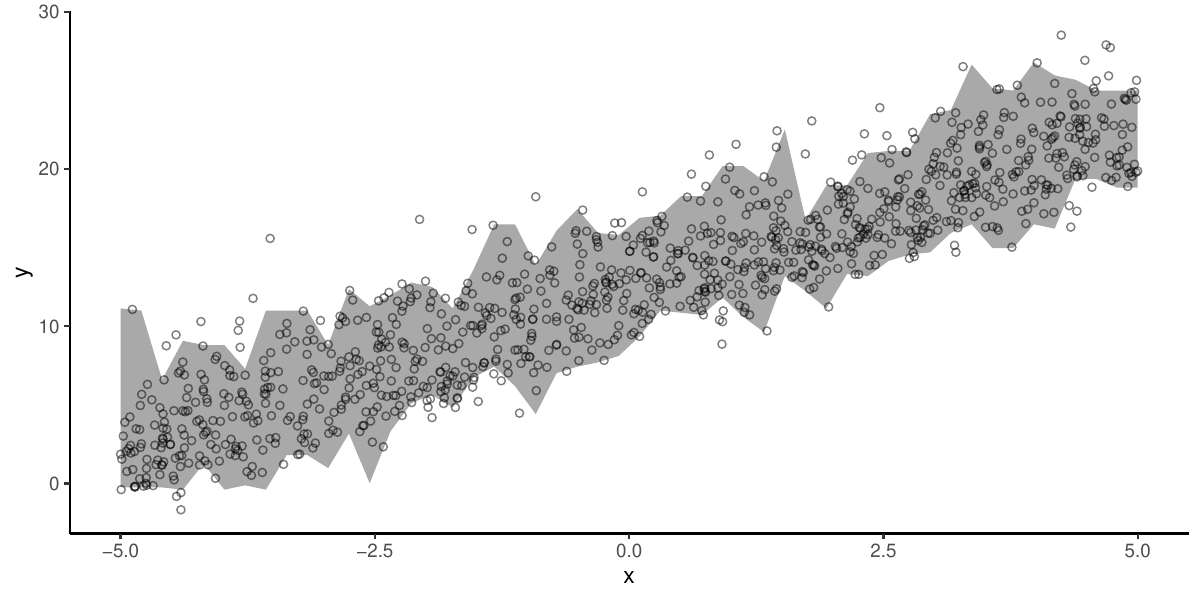}
\includegraphics[scale = 0.35]{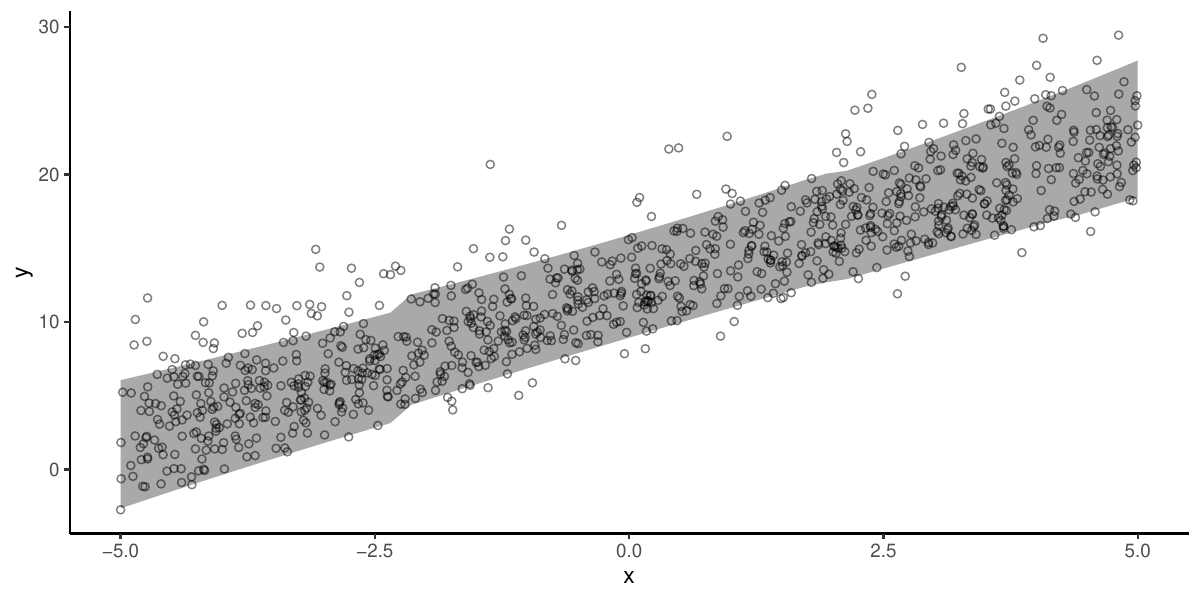}
\includegraphics[scale = 0.35]{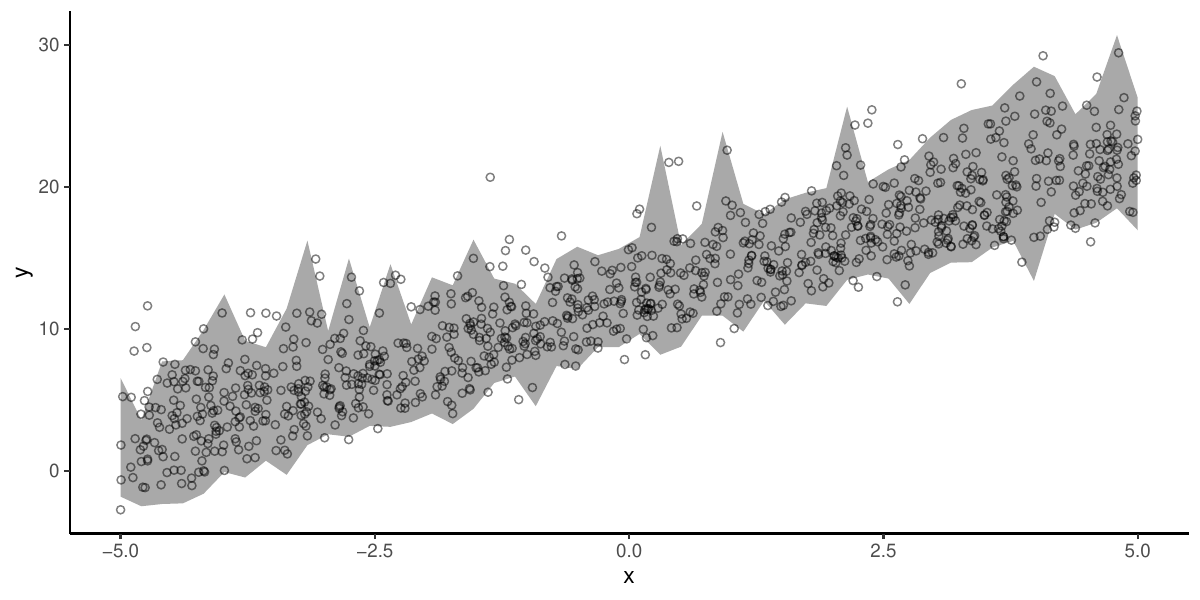}
\includegraphics[scale = 0.35]{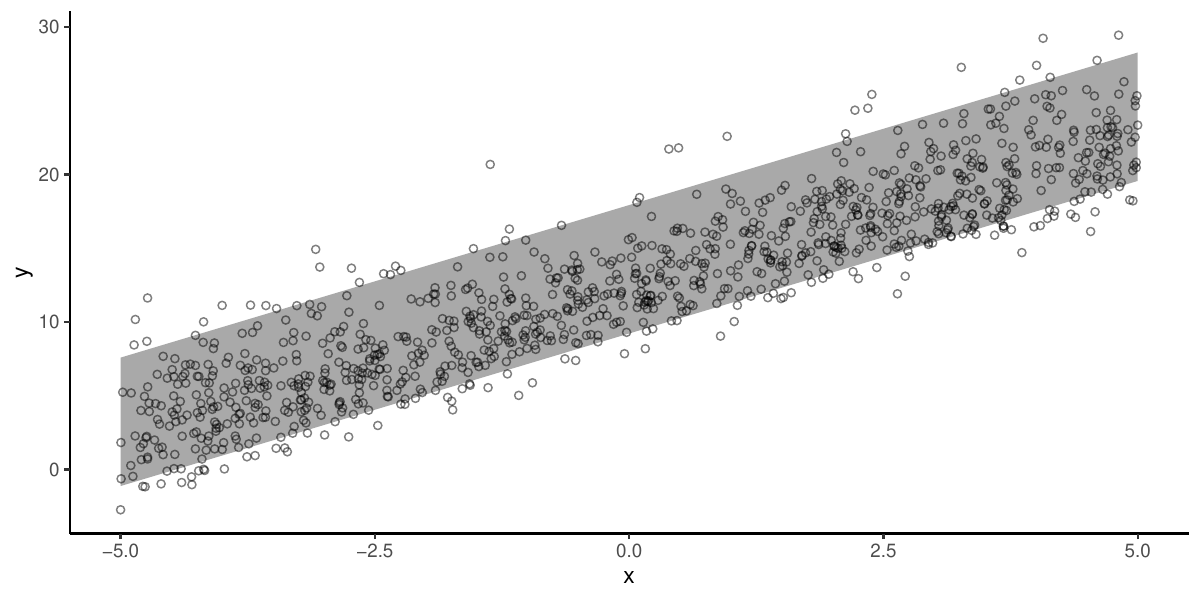}
\caption{Prediction Regions: Unimodal and Skewed. The shaded region is the prediction set from one simulation. From top left to right: HPD-split, CHR, DCP, CQR, and KDE-HPD}\label{fig:pred_regions2}
\end{figure}

\begin{figure}[H]
    \centering
\includegraphics[scale = 0.35]{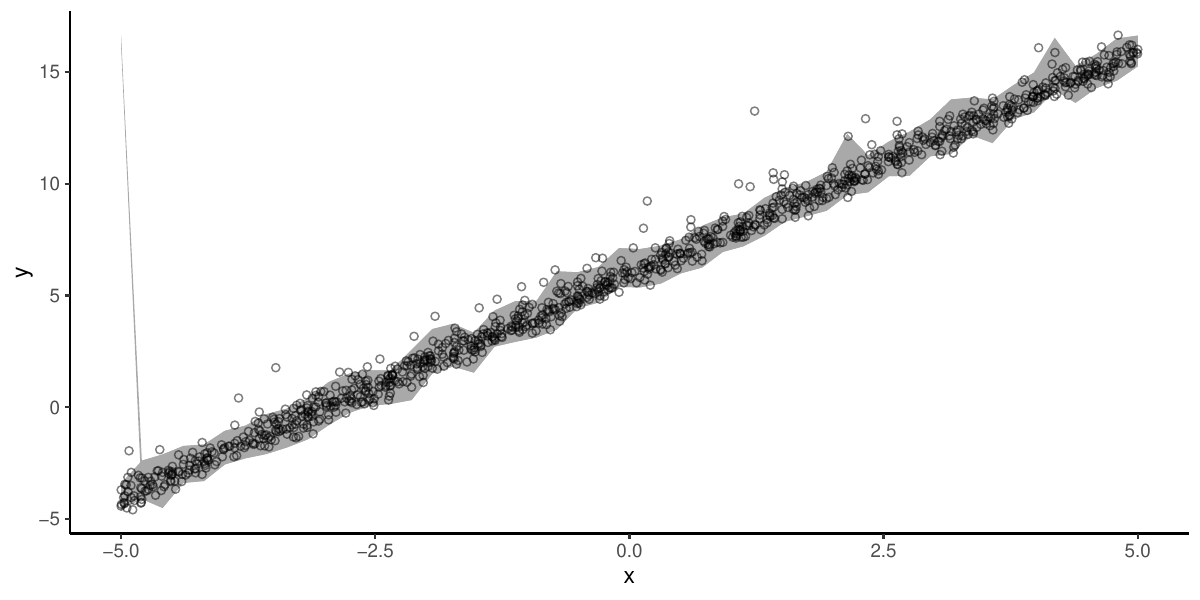}
\includegraphics[scale = 0.35]{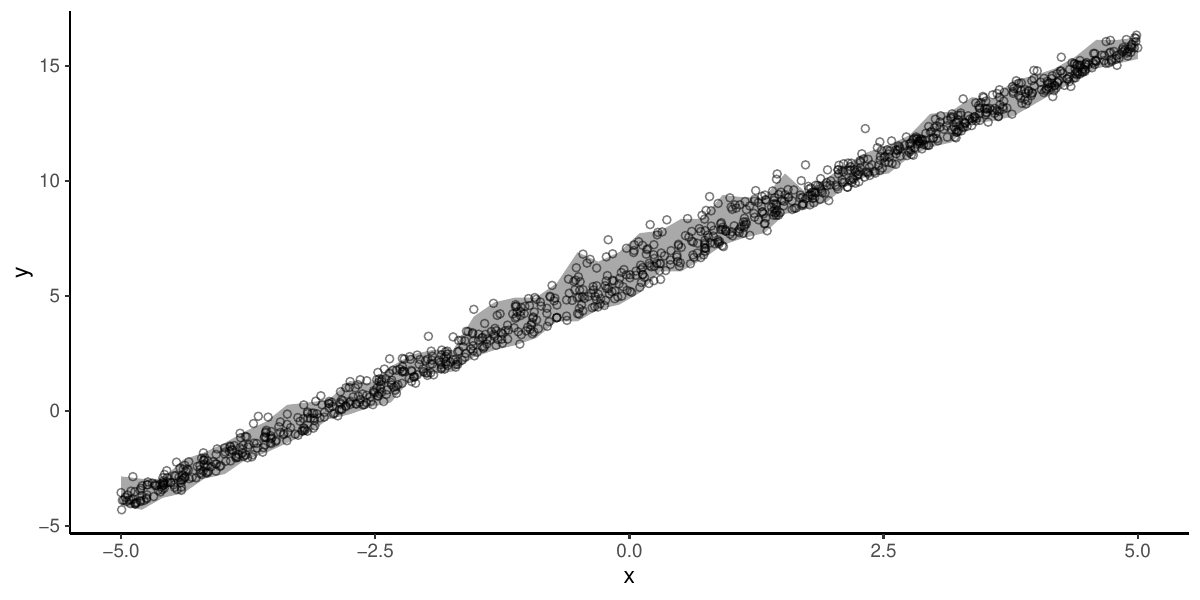}
\includegraphics[scale = 0.35]{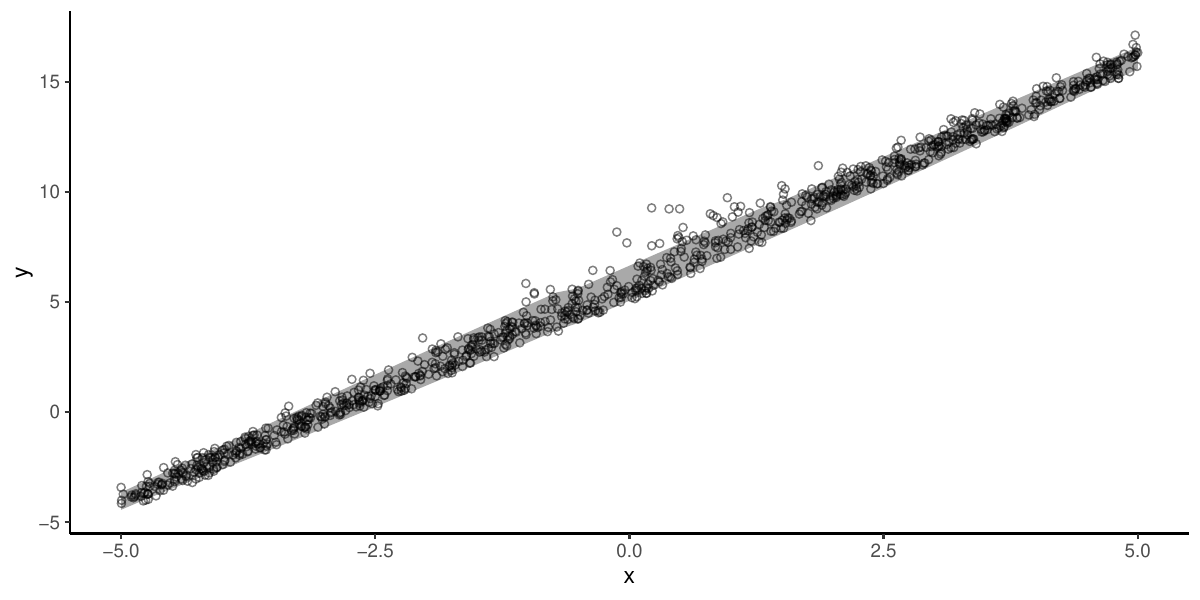}
\includegraphics[scale = 0.35]{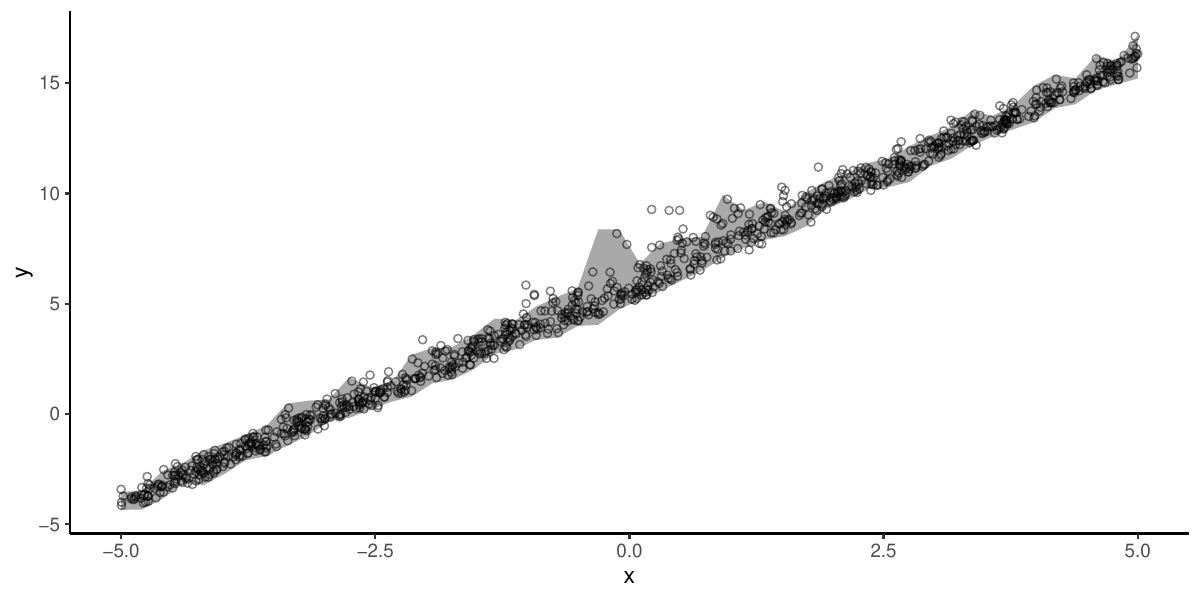}
\includegraphics[scale = 0.35]{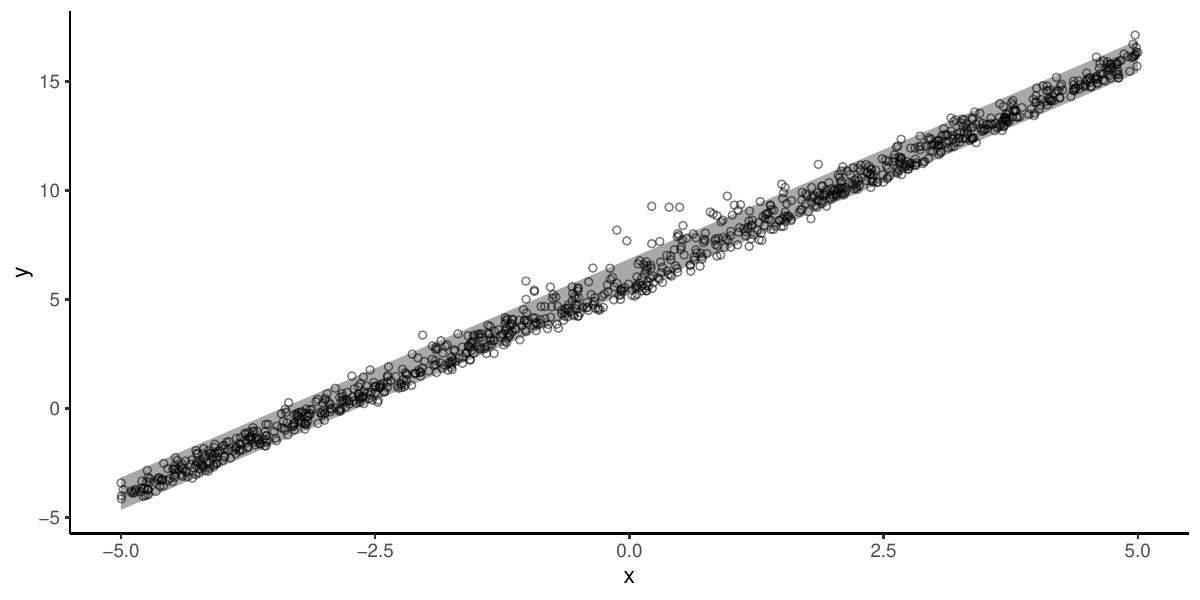}
\caption{Prediction Regions: Heteroskedastic. The shaded region is the prediction set from one simulation. From top left to right: HPD-split, CHR, DCP, CQR, and KDE-HPD}\label{fig:pred_regions4}
\end{figure}

\newpage

\section{KDE-HPD vs Parametric Prediction Sets}
\begin{figure}[ht]
    \centering
\includegraphics[scale = 0.5]{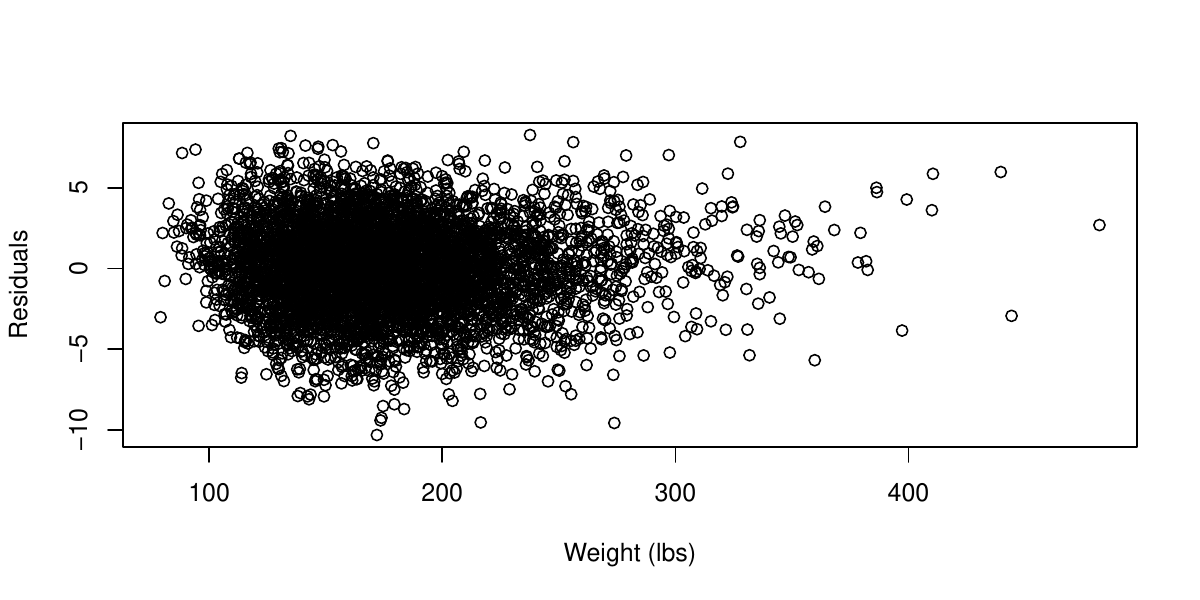}
\caption{Residuals from a linear regression vs a subject's weight}\label{fig:weight_resids}
\end{figure}

We compare KDE-HPD with Normal parametric prediction intervals on a portion of the NHANES 2005-2006 data set with height, weight, and gender of individuals \parencite{NHANES}. For both KDE-HPD and the Normal parametric prediction intervals, a linear regression model was used where the covariates were gender and weight and the response was height. We can see in~\cref{fig:weight_resids} that the residuals from this model are fairly homoskedastic. There were 5,107 observations in the data set. For KDE-HPD, 2,000 were used for training, 2,000 were used for calibration, and 1,107 were used for out of sample prediction. For the parametric approach, 4,000 were used for training the model and 1,107 were used for out of sample prediction. The data were randomly permuted 2,000 times. The average coverage, conditional coverage on gender, and average length are given in~\cref{tab:nhanes_results1}.

\begin{table}[H]
\begin{center}
    \begin{tabular}{|c|c|c|c|c|}
        \hline
         Approach &  Coverage &  Coverage (Male) & Coverage (Female) & Size \\
         \hline
        KDE-HPD & 0.899 & 0.886 & 0.911& 8.957 (0.003)\\
        Parametric & 0.903 & 0.900& 0.916& 8.995 (0.001) \\
         \hline
    \end{tabular}
    \caption{NHANES Comparison 1}
    \label{tab:nhanes_results1}
\end{center}
\end{table}
Slight heteroskedasticity can be seen for individuals who weigh more than 280 pounds in~\cref{fig:weight_resids}, so we looked at a second comparison that included a model for the heteroskedasticity for KDE-HPD to attempt to improve the conditional coverage. In this case, 1,000 observations were used to train the conditional mean model, 1,000 observations were used to train the model for heteroskedasticity (a linear model where the response was $|Y_i - \hat{g}(\vect{X}_i)|$ and the covariates were weight and gender), 2,000 observations were used in the calibration set, and 1,107 observations were used for out of sample prediction. The Normal parametric approach was the same as the first scenario. The results can be found in~\cref{tab:nhanes_results2}. Adding this model slightly improves the conditional coverage without increasing the interval lengths very much.

\begin{table}[H]
\begin{center}
    \begin{tabular}{|c|c|c|c|c|}
        \hline
         Approach &  Coverage &  Coverage (Male) & Coverage (Female) & Size \\
         \hline
        KDE-HPD & 0.900 & 0.889 & 0.910 & 9.003 (0.003)\\
        Parametric & 0.903 & 0.900 & 0.916& 8.995 (0.001) \\
         \hline
    \end{tabular}
    \caption{NHANES Comparison 2}
    \label{tab:nhanes_results2}
\end{center}
\end{table}

A linear regression of weight vs height is often used in introductory classes as a realistic example of linear regression with Normal errors. Looking at this application in both scenarios, we can see the prediction sets output by KDE-HPD have slightly smaller lengths than the Normal parametric prediction intervals, showing that KDE-HPD can give similar results to other methods when their ideal conditions are met. Adding a second model to help with the slight heteroskedasticity for KDE-HPD slightly improves the conditional coverage without sacrificing the length. It's clear that when we use good models with KDE-HPD, the results are very good.